\documentclass[aps,prx,twocolumn,floats,showpacs,superscriptaddress,nofootinbib]{revtex4-1}

\usepackage{feynmp}
\usepackage{graphicx,epsfig}
\usepackage{times}
\usepackage{graphics,dcolumn,bm,float}
\usepackage{amssymb,amsmath,rotate,color}
\usepackage[title,titletoc,toc]{appendix}
\usepackage{graphicx}
\usepackage{wrapfig}

\usepackage{lipsum}
\usepackage{mwe}

\usepackage[pagebackref=false,colorlinks,linkcolor=blue,citecolor=blue,urlcolor=magenta]{hyperref}

\usepackage[mathlines]{lineno}
\usepackage{hyperref}
\usepackage{breqn}
\usepackage{bbold}
\usepackage{pgfplots}
\usepackage{float}
\usepackage{tkz-euclide}
\usepackage{braket}
\usepackage{physics}
\usepackage[caption=false]{subfig}
\usepackage[export]{adjustbox}
\usepackage{tikz}
\usetikzlibrary{through,calc}
\usetikzlibrary{positioning}

\newcommand{\om}{\Omega}
\newcommand{\be}{\begin{equation}}
\newcommand{\ee}{\end{equation}}
\newcommand{\ep}{\epsilon}

\newcommand{\bearr}{\begin{eqnarray}}
\newcommand{\eearr}{\end{eqnarray}}
\newcommand{\nn}{\nonumber}

\newcommand{\bsq}{{\boldsymbol{q}}}
\newcommand{\bsk}{{\boldsymbol{k}}}
\newcommand{\pr}{\partial}
\newcommand{\bs}{\boldsymbol}
\DeclareMathOperator{\Lagr}{\mathcal{L}}

\begin{document}
\preprint{}
\title{Electrodynamics of tilted Dirac/Weyl materials: A unique platform for unusual surface plasmon polaritons}
\author{Z. Jalali-Mola}
\email{jalali@physics.sharif.edu}
\affiliation{
Department of Physics$,$ Sharif University of  Technology$,$ Tehran 11155-9161$,$ Iran
}

\author{S.A. Jafari}
\email{jafari@physics.sharif.edu}
\affiliation{
Department of Physics$,$ Sharif University of  Technology$,$ Tehran 11155-9161$,$ Iran
}

\date{\today}

\begin{abstract}
The electrodynamics of Weyl semimetals (WSMs) is an extension of Maxwell's theory where in addition to field strength 
tensor $F_{\mu\nu}$, an axion field enters the theory which is parameterized by a four-vector $b^\mu=(b_0,\bs b)$. 
In the tilted Weyl matter (TWM) an additional set of parameters $\bs\zeta=(\zeta_x,\zeta_y,\zeta_z)$ enter the theory
that can be encoded into the metric of the spacetime felt by electrons in TWM.
This allows an extension of Maxwell's electrodynamics that describes electric and magnetic fields in TWMs
and tilted Dirac material (TDM) when $b^\mu=0$.
The tilt parameter $\bs\zeta$ appearing as off-diagonal metric entries mixing
time and space components mingles $\bs E$ and $\bs B$ fields whereby modifies the inhomogeneous Maxwell's equations. 
As an example of the application of the electrodynamics of TWMs, we study the surface plasmon polariton (SPP) in these systems. 
The peculiarity of SPP on the surface of TWM or TDM is that it is not merely the propagation of electromagnetic modes on the 
surface of a conductor. It also describes the propagation of electromagnetic waves at the {\it interface of two different spacetime geometries}.
In the case of TDM, we find a characteristic dependence of SPP spectrum on the tilt parameter $\zeta$ which can be used
map $\zeta$ from SPP measurements. 
In the case of TWM, depending on whether the interface with vacuum supports a Fermi arc or not, and whether the propagation
direction is along the Fermi arc or transverse to it, we find many unusual spectral features for SPP modes.
These include: (1) surface plasmons with much higher frequency than bulk plasmon frequency, (2) soft SPP modes at short length scales,
(3) Tilt controlled SPP window beyond which SPP modes are unstable, (4) kink in the SPP dispersion, (5) uniform group velocity 
near the "horizon" ($\zeta=1$) and (6) possible negative group velocity. 
Our detailed study of the dependence of SPP spectra on the arrangements of three vectors $(\bs b,\bs q,\bs\zeta)$, the
first two of which are at our control, can be utilized to map the tilt characteristics and Fermi arc characteristics from SPP measurements. 
\end{abstract}



\maketitle
\narrowtext

\section{introduction}
Maxwell's electrodynamics describes propagation of electromagnetic waves in the vacuum.
Presence of matter is included by appropriate dielectric function $\epsilon(\bs q,\omega)$~\cite{JacksonBook} . 
In the case of dielectrics (ordinary insulators), the Maxwell theory is almost identical to 
that of vacuum, except for the replacement $\epsilon_0\to\epsilon_r$ of the dielectric constant. 
In a second class of materials known as topological insulators (TIs)~\cite{HasanMooreReview}, as well as in Weyl semimetals~\cite{FeslerAnnRev2017},
the Maxwell theory, in addition to the electromagnetic fields (encoded into the field strength 
tensor), will also contain a new pseudo scalar field called axion field, 
$\theta(\bs r,t)$~\cite{Wilczek1987,ZhangPhysToday,Tewari2013,Burkov2013,WittenZhang2013,Vazifeh2013,Hosur2013}. 
In the case of TIs, the spatial profile of the axion field is a step
function that drops to zero at the interface of the TI with vacuum. In the case of WSMs, the
axion field $\theta(\bs r,t)=-2b_0t+2\bs b.\bs r$ has a linear spacetime dependence which is
completely characterized by a four-vector $b^\mu=(b_0,\bs b)$~\cite{Tewari2013,Burkov2013,Ouellet2019}. 
In the case of time-reversal breaking
WSMs, the two Weyl nodes are at momenta $\pm \bs b$, and $2\bs b$ will be the momentum-space separation of 
right-handed (right-chirality) and left-handed Weyl fermions~\cite{Armitage2018,FeslerAnnRev2017}
WSMs and their associated Fermi arcs are experimentally realized in variety of 
systems~\cite{Lv2015,Lu2015Sience,Xu2015,Xu2016,Huang2016,HasanAnnualRev2017}. 
Similarly $b_0$ is a possible energy separation of the Weyl nodes. 
In the case of Dirac materials, the separation $b^\mu=0$ and the two copies of Weyl node coincide. 
The above axion field has intimate connection with the chiral anomaly~\cite{NielsenNinomiya1983,Zyuzin2012,Kharzeev2014,Pesin2015} 
that stems from the transformation of the measure of path integral under chiral rotation~\cite{Fujikawa1979,ZeeQFTBook,Peskin}
and implies the particle number with definite chirality is not conserved. 
The electrodynamics of such systems contains an additional $\theta(\bs r,t) \bs E.\bs B$ term in
their Lagrangian~\cite{Wilczek1987,Vazifeh2013,Burkov2013,Zyuzin2012,Tewari2013,ZhangPhysToday,FeslerAnnRev2017,Brazovskii2019}.

Let us now think of the possible deformations of the cone-shaped dispersion in Dirac and Weyl materials.
One interesting deformation is to tilt the Weyl cones to obtain TWMs~\cite{Soluyanov2015,Deng2016,Yun2016,Zyuzin2016,Goerbig2016PRL}
or TDMs~\cite{Katayama2008,Kajita2010,Goerbig2008,Goerbig2009,Suzumura2014,Goerbig2014,Noh2017,Tohyama2009,TohidBorophene}.
The tilt can be in either of three space directions and hence is characterized by a triple of tilt parameters $\bs\zeta=(\zeta_x,\zeta_y,\zeta_z)$~\cite{SaharCovariance}.
When the magnitude $\zeta$ of this vector satisfies $\zeta<1$ ($\zeta>1$) it corresponds to type-I (type-II) Dirac/Weyl 
material~\cite{Zhang2014,Weng2015,Wan2011,Soluyanov2015,Deng2016,Huang2016}. Type-I (type-II) cones have closed (opened) Fermi surfaces
which are separated by a Lifshitz transition~\cite{Volovik2018}. 
In a minimal model of WSM containing two Weyl nodes, the two cones will be tilted
in opposite directions. By setting $b^\mu=0$ the axion physics disappears and
the TWM reduces to the corresponding TDM. 
The tilt leave its signatures in quantum transport~\cite{Bergholtz2015}
non-universal anomalous Hall effect~\cite{Zyuzin2016}, non-zero anomalous Nernts effect (despite linear dispersion)~\cite{Ferreiros2017}, 
squeezed Landau levels~\cite{Yu2016}, kink in the plasmon dispersion~\cite{SaharTilt1,SaharTilt2}, amplification of magnetic fields~\cite{JafariEB}.
Within the geometric language, the tilt deformation can be encoded into a spacetime metric~\cite{Volovik2016}.
From this perspective, TWMs realize a Lorentz violating quantum filed theory~\cite{Yan2107}. Nevertheless, 
still a large enough symmetry group is left which is a continuous deformation of the Lorentz group~\cite{JafariEB}.  
As will be detailed in next section, the metric $g_{\mu\nu}$ of the deformed Minkowski spacetime in tilted cone materials is a 
continuous deformation of the metric $\eta_{\mu\nu}={\rm diag}(-1,1,1,1)$~\cite{SaharCovariance,JafariEB}.
Then the Lifshitz transition from $\zeta<1$ to $\zeta>1$ side in a geometric language will be identified with a black-hole horizon~\cite{Ryder,Padmanabhan}. 
It has been suggested that the value of tilt in a certain structure of borophene sheet~\cite{Zhou2014,Lopez2016,Tarun2017} can 
be tuned with perpendicular displacement fields~\cite{TohidBorophene}. 

Departing from the point that the tilt deformation in TWMs and TDMs can be encoded into the spacetime
metric, it would be natural to explore how does the modified metric affect the Maxwell theory?
In this paper, we will derive the Maxwell theory in a non-Minkowski background metric of TWMs.
Setting the axion field parameter $b^\mu=0$ (i.e. dropping the axion field) gives the electrodynamics
theory for TDMs. It turns out that encoding the electrodynamics response of the system into a 
metric provides a mathematically neat way to incorporate the tilt parameter $\bs\zeta$
into the electrodynamic response of the system~\cite{SaharCovariance}.

As an example we will consider the propagation of the electromagnetic waves on the interface of 
vacuum (with $\bs\zeta=0$ and $b^\mu=0$) with a TWM (both $\bs\zeta$ and $b^\mu$ non-zero) or TDM (only $\bs\zeta$ non-zero). 
These modes are called surface plasmon polaritons (SPP) which are the electromagnetic excitations of the electron systems trapped on the surface of 
two media. 
SPPs have potential technological application. The SPP resonance comes from the interaction of electromagnetic waves with the longitudinal 
oscillation of charge density on the interface of system with either vacuum or another dielectric.
These modes decay evanescently into the bulk of both media in which the dielectric constants have opposite 
signs~\cite{Ritchie1957,Boardman-Book,Raether-Book,Homola1999,sarid-challener-2010,Zhang2012}. 
Hence the SPP modes are localized on the interface and are employed in control and manipulation of light on the nanometre scale 
(smaller than light wavelength)~\cite{Zhang2012,ZayatsReview} which has applications in subwavelength 
photonic~\cite{Barnes2003,Maier2003}, chemical and bio-sensors~\cite{Homola1999,Nemova2006,Anker2008},  
data storage~\cite{Hermann2001,Zijlstra2009}, solar cells~\cite{Stenzel1995,Westphalen2000} and also in near field optics~\cite{Kim1995,Pendry2000}.
The above list is a selection of applications of the SPP modes otherwise, the list is much wider and the reader is referred to 
literature of the field~\cite{ZayatsReview}. Given the enormous technological applications of SPPs, it would be nice
to have a new class of systems where further control/filtering on the SPP modes can be achieved.

As will be shown in this paper, the interface of TWMs or TDMs with vacuum or a dielectric is quite 
a unique interface. In one side of the interface -- as far as the propagation of electromagnetic waves
is concerned -- the spacetime is non-Minkowski, while in the other side the spacetime is Minkowski. 
Therefore the resulting surface modes reflect characteristics of the changes that occur because of 
the abrupt change in the metric of the spacetime across the interface. From this perspective, 
{\it TWMs and TDMs furnish a unique surface across which the geometry changes}. The SPP modes localized into
this surface are subject of the present work. 

To advertise some of the features of SPP modes in TWMs and TDMs:  (i) On the surface of TDMs, the saturation 
frequency $\Omega_s$ corresponding to surface plasmons~\cite{Ritchie1957} 
will be heavily affected by the tilt parameter $\zeta$. (ii) In the case of TWMs, 
a SPP-forbidden region arises from strong enough $|\bs b|$ axion parameter within which 
the SPP and bulk plasmons hybridize. The tilt parameter is encoded in the spacetime metric
will have interesting effects on this region. (iii) There can be circumstances, 
where SPP modes with frequencies larger than the bulk plasmon frequency $\Omega_p$ will be possible.
This is in sharp contrast to the surface of ordinary conductors where  
the dielectric constant $1-\Omega_p^2/\omega^2$ of the Drude theory is {\em positive} for $\omega>\Omega_p$
and therefore no SPP modes exist. Therefore the already wide range $0<\omega<\Omega_s$ of the SPP modes~\cite{Kolb1982,Agranovich-book}
can be further extended. Such an ultra-high energy SPP can be detected by scanning near field technology~\cite{Kolb1982,ZayatsReview,Agranovich-book}. 
(iv) Tilt can induce a down bending of the SPP modes
which gives rise to {\it negative group velocity} for SPP modes. 
(v) For certain facets of TWMs, in the "near horizon" limit $\zeta\to 1$, 
there can be soft SPP modes near $q\to\infty$, namely at low wave-length (atomic scales).

This paper is organized as follows. In section~\ref{sec2} we formulate the
Maxwell theory with an additional $\bs\zeta$ vector. 
In section~\ref{sec3} we set the axion parameters zero, and focus on the role of $\bs\zeta$ on the
SPP modes of a TDM. In section~\ref{sec4} we study SPP modes in TWMs with $b^\mu=(0,{\bs b})$ in various situations. 
In section~\ref{sec5} we study the SPP modes for TWM with $b^\mu=(b_0,\bs 0)$. 
We end the paper in section~\ref{sec6} with a summary and outlook.

\section{Electrodynamics of tilted Dirac/Weyl fermions}{\label{sec2}}
The minimal model for a single Weyl node of a fixed chirality is given by~\cite{Goerbig2008,Goerbig2009,Feng2017},
\begin{equation}
  H(k)=\hbar( v_F \bs{\sigma}.\bsk+\bs v_t.\bsk \sigma_0),
  \label{Hamiltonian.eqn}
\end{equation}
where $\bs k$ is the momentum measured from the Weyl node, and $\bs\sigma=(\sigma_1,\sigma_2,\sigma_3)$ denotes Pauli matrices. 
The $\sigma_0$ is the $2\times 2$ identity matrix $\bs 1$. 
Here the first term denotes to the upright Weyl Hamiltonian and the second term gives the tilt of dispersion relation. 
The Fermi velocity in the isotropic case is $v_F$  and the tilt characteristic of Weyl fermions arises from velocity vector $\bs v_t$ 
in which each component of $\bs v_t$ denotes to the amount of tilt along the corresponding direction. $v_F$ characterizes the 
solid angle (future "light" cone) subtended by the Dirac dispersion relation, while the $\bs v_t$ characterizes the magnitude
and direction of the tilt. It is convenient to define a dimensionless set of parameters $\zeta_i=v_{t,i}/v_F$. 
Therefore the above $\bs\zeta$ completely characterizes the tilt of the Weyl cone. 
The eigenvalues of the above Hamiltonian give the following dispersion relation
\be
   E_s(\bs k)=\hbar v_F(s|\bs k|+\bs \zeta.\bs k),
   \label{dispersion.eqn}
\ee
where the sign $s$ denotes upper ($+$) and lower ($-$) branches of the cone. 
Hence $0 \le \zeta \le 1$ denotes to the Type-I Weyl fermions which also includes upright Weyl semimetals and 
$\zeta>1$ represents  type-II in which the black hole horizon place at the boundary of these two groups with $\zeta=1$. 

By crystal symmetries, there has to be at least one more Weyl node of the above form. If the time-reversal and
inversion (parity) are respected, one will have a 3+1 dimensional Dirac theory where another copy of the
Hamiltonian~\eqref{Hamiltonian.eqn} with opposite chirality is superimposed on it~\cite{ZeeQFTBook,Peskin,Yang2014,WehlingReview}. 
If either of the time-reversal or inversion symmetry is broken, the two Weyl nodes will be separated
in momentum space, and the resulting material will be a Weyl semimetal~\cite{FeslerAnnRev2017} the first example of which was realized
in TaAs~\cite{Lu2015Sience,Lv2015}. 
The tilt direction in the other valley is opposite to first one ($-\bs v_t$)~\cite{TohidBorophene}.
The Weyl semimetals are characterized by a four-vector $b^\mu=(b_0,{\bs b})$
which corresponds to the energy-momentum separation of the two Weyl nodes. 
$\bs b$ breaks the time reversal symmetry~\cite{Zyuzin2012,Tewari2013,Hosur2013}

If we ignore the tilt term in Eq.~\eqref{Hamiltonian.eqn} (i.e. set $\bs v_t=0$)  we will have the upright Weyl Hamiltonian
with $\zeta=0$. From geometrical point of view, the Weyl equation is a Lorentz invariant in 3+1 dimensional spacetime.
This allows to recast the dispersion relation in terms of the Minkowski metric $\eta_{\mu \nu}={\rm diag}(-1,1,1,1)$ in a compact
and invariant form  $\eta^{\mu \nu} k_\mu k_\nu=0$ with $k_{\mu}=(E/\hbar v_F,\bsk)$. For massive (gapped) excitations, the
right-hand side of this equation will be determined by the mass. The presence of tilting breaks the Lorentz symmetry 
into a smaller (but still rich enough) symmetry~\cite{JafariEB}. The resulting new spacetime will be described
by a deformation of the Minkowski metric as
\be
g_{\mu \nu}=
\begin{bmatrix}
-1+\zeta^2	&	 -\zeta_x	&	-\zeta_y	&	-\zeta_z \\
 -\zeta_x	&	1	&	0	&	0 \\
 -\zeta_y	&	0	&	1	&	0\\
 -\zeta_z	&	0	&	0	&	1\\
\end{bmatrix},
\label{tmetric.eqn}
\ee
in which $\zeta^2=\zeta_x^2+\zeta_y^2+\zeta_z^2$ is the total magnitude of the tilt in 3D. 
With the above metric, one can write the dispersion relation Eq.~\eqref{dispersion.eqn} as
\be
    g_{\mu\nu}k^\mu k^\nu=0. 
\ee
Indeed with the above metric, one can express the polarization tensor for the undoped
Dirac cone in a covariant form~\cite{SaharCovariance}. This implies that as far as the
electromagnetic response of free electrons in tilted Dirac/Weyl system is concerned, the spacetime
for the electrons looks as Eq.~\eqref{tmetric.eqn}. The natural question would be, 
how are the Maxwell equations governing the electromagnetic fields themselves will be modified in such a spacetime?

The field strength tensor in the Maxwell theory is given by $F_{\mu\nu}=\partial_\mu A_\nu-\partial_\nu A_\mu$ in which 
$A_\mu=(-\phi,\bs A)$ and $\partial_\mu=(c^{-1}\partial/\partial t,\bs \nabla)$. 
In the deformed Minkowski space given by metric~\eqref{tmetric.eqn}, the contravariant components of $F$ 
will be given by, $F^{\mu\nu}=g^{\mu\alpha}g^{\nu\beta} F_{\alpha\beta}$, which turns out to be
\be
F^{\mu \nu}=
\begin{bmatrix}
0	&	 \bar{E}_x	&	 \bar{E}_y	&	 \bar{E}_z \\
  -\bar{E}_x	&	0	&	\bar{B}_z	&	-\bar{B}_y \\
 - \bar{E}_y	&	-\bar{B}_z	&	0	&	\bar{B}_x\\
  -\bar{E}_z	&	\bar{B}_y	&	-\bar{B}_x	&	0\\
\end{bmatrix},
\label{Fup.eqn}
\ee 
with
\bearr
&&\bar{\bs E}=\bs E+ \bs\zeta \times \bs B, \nn\\
&&\bar{\bs B}=\bs B(1-\zeta^2)+\bs\zeta (\bs\zeta.\bs B)+\bs\zeta\times\bs E,
\eearr
where the bar notation on electric and magnetic field strength emphasizes the role of tilt parameter $\bs\zeta$. 
It is pleasant to note that the above relations reduce to textbook expressions for $F^{\mu\nu}$ components
in the $\bs\zeta=0$ limit~\cite{Ryder}. 

Before writing down the Lagrangian, it is useful to discuss the duality in the TWM. 
In the normal Minkowski spacetime if one defines the Lev-Civit\'a symbol $\ep^{\mu\nu\alpha\beta}$ to have
value $1$ ($-1$) when $(\mu\nu\alpha\beta)$ is even (odd) permutations of $(0123)$, and zero otherwise~\cite{Ryder}.
This helps to define the dual field strength tensor defined by $\tilde F^{\mu\nu}=(1/2)\ep^{\mu\nu\alpha\beta}F_{\alpha\beta}$.
In Minkowski spacetime one has the following duality relation
\be
   (\bs E\to\bs B,\bs B \to -\bs E) \equiv (F^{\mu\nu}\to \tilde F^{\mu\nu}).
   \label{duality.eqn}
\ee
Now in the TWM case, the same definition for Levi-Civit\'a symbol implies the following duality relation
\be
   (\bar{\bs E}\to{\bs B},~~\bar{\bs B} \to -{\bs E}) \equiv (F^{\mu\nu} \to \tilde F^{\mu\nu}).
   \label{tduality.eqn}
\ee
But since in the $\bs\zeta\to 0$ limit, $(\bar{\bs E},\bar{\bs B})\to(\bs E,\bs B)$, the duality 
relation~\eqref{tduality.eqn} of TWM reduces to the duality relation~\eqref{duality.eqn} of the Minkowski spacetime. 
As can be seen, the tilt parameter $\bs\zeta$ explicitly enters the duality relation~\eqref{tduality.eqn}. 

The Lagrangian of the electromagnetic (EM) fields in the tilted Dirac materials is given by a $-F_{\mu\nu}F^{\mu\nu}/4$~\cite{JacksonBook}
where $F^{\mu\nu}$ is given by Eq.~\eqref{Fup.eqn}. In the case of WSMs with a given $b^\mu$ that characterizes an axion field
 $\theta(\bs r,t)$~ the total Lagrangian will become~\cite{Zyuzin2012,Burkov2013,Tewari2013,Fujikawa1979},
\be
\Lagr_{em}=-\frac{1}{4} F_{\mu\nu}  F^{\mu\nu}-\frac{1}{4} g_a \theta(\bs r,t) \tilde F^{\mu\nu}F_{\mu\nu}+ j^\mu A_\mu,
\label{Lem.eqn}
\ee
where $g_a=e^2/4\pi^2 \hbar c$  is the coupling constant between axion and polaritons~\cite{Hofmann2016}
and the last term is the minimal coupling with the external source $j^\mu=(c \rho,\bs j)$.
As pointed out, the second term appears in effective Maxwell theory for topological insulator and WSMs~\cite{ZhangPhysToday,HasanMooreReview}.
This term actually arises from integration of matter field which leaves behind the above so called
theta term~\cite{Burkov2013,Fujikawa1979,Peskin}. The essential feature of the above theta term is that
it is a geometric invariant and its form is independent of whether the metric is $\eta_{\mu\nu}$ or the metric $g_{\mu\nu}$ of Eq.~\eqref{tmetric.eqn}~\cite{Ryder}.
Therefore the form of the theta term in both tilted and non-tilted Weyl semi-metals is the same and is proportional to
$\tilde F F \sim \bs E.\bs B$. It is important to note that the duality relation~\eqref{tduality.eqn} can be used to 
obtain $\tilde F$ from $F$ in TWMs. In doing so, eventually the $\bs E$ and $\bs B$ are mapped to those without bar, 
and therefore the above $\bs E.\bs B$ form is the same for both Minkowski spacetime and TWMs. This invariance is expected from the very 
nature of topological terms as they do not depend on the coordinates and metric~\cite{CarrolBook}. 
At the end, all the changes to standard Maxwell equations in tilted Weyl cone medium is encoded
in the first term that includes contravariant components of $F$ as given by Eq.~\eqref{Fup.eqn}. 

We are now ready to discuss the Maxwell's equations in TWM. Let us start with the homogeneous equations, $\partial_\mu \tilde F^{\mu\nu}=0$~\cite{Ryder}.
Since the $\tilde F$ as discussed above is solely determined from $F_{\mu\nu}$, it will not contain any effect  of the tilt. 
Therefore the homogeneous Maxwell equations will remain intact:
\bearr
   && \bs\nabla \times \bs E=\frac{1}{c}\frac{\partial \bs B}{\partial t},
   \label{curlE.eqn}\\
   && \bs\nabla \cdot \bs B =0.
   \label{divB.eqn}
\eearr
However, the inhomogeneous Maxwell equations $\partial_\nu F^{\mu\nu}=(4\pi/c)j^\mu$~\cite{Ryder}
will be strikingly modified as a result of non Minkowski metric,~\eqref{tmetric.eqn}.
The most generic form of Maxwell's equation including both the tilt $\bs\zeta$ and the axion effect
$\theta$ can be obtained as follows:
\bearr
&& 4\pi \rho= \bs\nabla.\bs E-\bs\zeta.\bs\nabla \times \bs B-4 \pi g_a(\bs\nabla\theta).\bs B, \label{rho.eqn}\\
&& \frac{4\pi}{c} \bs j=-\frac{1}{c} \frac{\pr \bs E}{\pr t} -\frac{1}{c}\bs \zeta \times \frac{\pr \bs B}{\pr t}+\bs\nabla\times(\bs\zeta \times \bs E)+\label{j.eqn}\\
&&(1-\zeta^2) \bs\nabla \times \bs B-(\bs\zeta\times \bs \nabla)(\bs\zeta.\bs B)+\frac{4\pi}{c} g_a\dot{\theta} \bs B+
\nn\\&&
4\pi g_a \bs\nabla\theta\times\bs E.\nn
\eearr

From above relations it can be  easily seen that the metric changes the electromagnetic response 
of the bulk material in the way that the displacement field will be given by
\bearr
\bs D=&&(\ep_\infty+\frac{4\pi i \sigma}{\omega}) \bs E+ \bs\zeta \times \bs B-\frac{i c}{\omega} \bs\nabla \times \bs \zeta\times \bs E+\nn\\&&
\frac{i c}{\omega} \zeta^2 \bs\nabla \times \bs B+\frac{i c}{\omega} (\bs\zeta \times \bs \nabla)(\bs\zeta.\bs B)+\frac{4\pi i }{\omega} g_a \dot{\theta}\bs B+\nn\\&&
\frac{4\pi i c}{\omega} g_a (\bs\nabla\theta)\times E,
\label{disp1.eqn}
\eearr
in which  the first term $(\ep_\infty+\frac{4\pi i \sigma}{\omega}) $ represents the plasmon oscillation of bulk matter 
which in the nearly free electron approximation in a dielectric environment with background dielectric
constant $\ep_r$ is equal $\ep(\omega)=\ep_r(1-\Omega_p^2/\omega^2)$. 
In undoped Weyl nodes where the density of electrons (with respect to charge neutrality point) is zero, the $\Omega_p$ will become
zero and the above dielectric function will become a constant $\ep_r$. In the high frequency limit, $\omega\to\infty$, irrespective
of value of $\Omega_p$, the dielectric function will reduce to $\ep_r=\ep_\infty$. 
Using Eq.~\eqref{curlE.eqn} in the right hand side Eq.~\eqref{disp1.eqn} , the second and third terms cancel each other and displacement 
acquires the final form,
\bearr
\bs D=&&(\ep_\infty+\frac{4\pi i \sigma}{\omega}) \bs E+
\frac{i c}{\omega} \zeta^2 \bs\nabla \times \bs B+\frac{i c}{\omega} (\bs\zeta \times \bs \nabla)(\bs\zeta.\bs B)+\nn\\&&
\frac{i }{\omega} g_a \dot{\theta}\bs B+
\frac{i c}{\omega} g_a (\bs\nabla\theta)\times E.
\label{dielectric.eqn}
\eearr
Combining the four Maxwell's equations in the presence of tilted Weyl matter 
gives, 
\be
\bs \nabla \times \bs \nabla \times \bs E+\frac{1}{c^{2}}\frac{\pr^2 }{\pr t^2}\bs D=0
\label{Mtensor.eqn}
\ee
where the information of the specific tilted Weyl material is encoded in the displacement Eq.~\eqref{dielectric.eqn}. 
Note that in TWM, for any Weyl node of tilt $\bs\zeta$, there exists a corresponding node of
tilt $-\bs\zeta$. As long as there are not atomically sharp scatterers to mix the two nodes,
the two nodes can be treated as two independent blocks, and therefore their dielectric response
will be additive. The cancellation leading to Eq.~\eqref{dielectric.eqn} has already canceled
the terms that are odd in $\bs\zeta$. Therefore the presence of the second Weyl node with opposite
tilt does not alter Eq.~\eqref{dielectric.eqn}, except for multiplying the displacement field
by a degeneracy factor of two.

From the perspective of Maxwell's equations, the ordinary dielectrics (band insulators) behave
like the vacuum, except that the dielectric constant of vacuum, $\epsilon_0$ is replaced by the
dielectric constant $\epsilon_r$ or dielectric function $\epsilon(\bs q,\omega)$~\cite{JacksonBook}. Topological insulators will be
a distinct class of dielectrics where the Maxwell equation will be modified by utilizing the second term
in Eq.~\eqref{Lem.eqn}, the theta term~\cite{ZhangPhysToday} that contracts $\tilde F$ with $F$. 
Now the tilted Dirac/Weyl matter presents a third distinct class of materials with respect to their electromagnetic properties
where an additional parameter $\bs\zeta$ appears in the Maxwell theory as in Eqs.~\eqref{rho.eqn},~\eqref{j.eqn},
and~\eqref{dielectric.eqn}.
The standard model of particle physics does not have the $\bs\zeta$ term. Because this tilt term 
arises from the underlying lattice~\cite{TohidBorophene,JafariEB,SaharCovariance}. Indeed the so called
non-symmorphic lattice is able to provide low-energy electronic excitations that have no counterpart
in the standard model of particle physics~\cite{BernevigNexus}. From this perspective, the tilted
Dirac/Weyl matter furnishes a unique form of electromagnetism which has no counterpart in the
standard model of particle physics.

\begin{figure}[t]
   \centerline{\includegraphics[width = .4\textwidth] {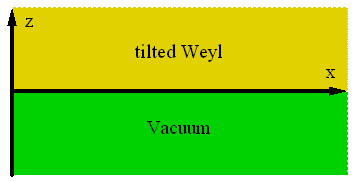}}
   \caption{(Color online) Schematic illustration of the geometry of the interface between
   tilted Weyl matter ($z>0$ region) and vacuum ($z<0$ region).
   The exponential term in TWM is  $\exp(-\gamma_1 z)$ and in the vacuum side is $\exp(\gamma_0 z)$. }
   \label{fig1-slab.fig}
\end{figure}

As an example of this unique electromagnetism theory, 
in the rest of this paper we focus on the SPP excitations at the interface between tilted Weyl semimetals and vacuum (it can also be replaced
by another normal dielectric). 
For this purpose, we consider the situation depicted in Fig.~\ref{fig1-slab.fig} where the surface of the tilted Weyl matter is located in the $xy$ plane 
in such a way that the bulk of the tilted Weyl matter occupies the $z>0$ region while the vacuum is in the $z<0$ side. 
We will be searching for solutions of Eq.~\eqref{Mtensor.eqn} subject to constituent equation~\eqref{dielectric.eqn} of the tilted Weyl matter.
Since the SPP solutions are localized in the surface, their electromagnetic fields exhibit exponential decay in the direction perpendicular to the surface. 
Hence the SPP solutions we will be seeking  have electric field profiles which propagate in $xy$ plane and decay as a function of distance $|z|$
way from the interface as
\be
\bs E=\bs E_0 e^{i\bs q.\bs r-\omega t} e^{-\gamma |z|},
\label{electric-t-c.eqn}
\ee
where $\bsq=(q_x,q_y)$. In this paper we will use the convention $\bs q=q \hat{\mathbf q}$ where $\hat{\mathbf q}$
denotes the direction of $\bs q$. 
Note that the decay strength $\gamma$ differs in the two sides of interface. 
For definiteness we consider one side to be vacuum as in Fig.~\ref{fig1-slab.fig} and denote the corresponding
index by $0$ and the decay constant in the tilted Weyl matter by $1$. 
Furthermore to obtain the SPP dispersion we need to use the boundary condition $E_{0x(y)}=E_{1x(y)}$ and $D_{0,z}=D_{1,z}$ 
and $\bs B_0=\bs B_1$ (for non-magnet material). 
Note that the same equation can be used to derive the bulk plasmon as well. 
The only difference will be that as a result of propagation of plasmon in bulk, 
the decaying part in the Eq.~\eqref{electric-t-c.eqn} will be replaced by a 
propagating wave.

The effect of axion field parameterized by $b^\mu$ on the SPP modes has already been
addressed in the existing literature~\cite{Hofmann2016,Polini2015,Tamaya2019}. In this work, we
would like to understand the role of $\bs\zeta$ and the joint effect of $\bs\zeta$ and $b^\mu$ together
and their interplay with each other. In the following section we first set $b^\mu=0$ and consider
a tilted Dirac matter. Next, we consider a non-zero $b^\mu$ relevant to a Weyl semimetal, and 
investigate the interplay between tilt and axionic aspects.

\section{Tilted Dirac matter}{\label{sec3}}
Setting $b^\mu=0$ amounts to placing the two Weyl nodes on top of each other. Therefore the TWM will
actually become a TDM. In the absence of $b^\mu=0$ vector, and having fixed the direction of the surface normal to be
along the $z$ axis, there will only remain two vectors in the theory, namely the $\bs q$ and the tilt parameter $\bs\zeta$. 
Therefore the relative direction of momentum on the surface ($xy$ plane in Fig.~\ref{fig1-slab.fig}) 
of tilted Weyl semimetals and tilt direction on the SPP dispersion should be considered. 

\begin{figure}[t]
   \centerline{\includegraphics[width = .4\textwidth] {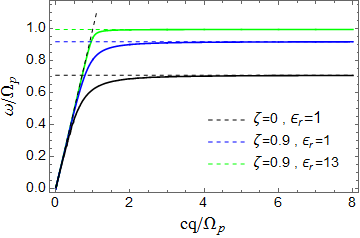}}
   \caption{(Color online) The SPP dispersion in the TDM (i.e. $b^\mu=0$) for $\bsq ||\bs\zeta$. 
   The black line is SPP for upright Dirac material with $\zeta=0$ and $\ep_r=1$. 
   Blue and green solid lines are SPP for strong tilt $\zeta=0.9$ with two different
   background dielectric constants $\ep_r=1,13$, respectively. The dashed lines
   represent the asymptotes corresponding to $q\to 0$ and $q\to\infty$. 
   }
   \label{fig2-xx0.fig}
\end{figure}

Let us start with the case where these two vectors are parallel. The rotational freedom in the $xy$ plane can be used to choose them 
as $\bs q=q\hat x$ and $\bs\zeta=\zeta \hat x$ both lying in the surface of TDM. 
The wave Eq.~\eqref{Mtensor.eqn} will become
\be
\begin{bmatrix}
	 -\gamma_1^2\lambda^2	&	 0	&	-iq\gamma_1\lambda^2 \\
  	 0	&	q^2\lambda^2-\gamma_1^2	&	0\\
     -iq\gamma_1\lambda^2	&		0	&	q^2 \lambda^2\\
\end{bmatrix}
\bs E
-\frac{\omega^2}{c^{2}} \ep_1(\omega)  \bs E=0,
\label{tensor-xx.eqn}
\ee
where $\lambda^2=(1-\zeta^2)$,  $\bs E^T=(E_x,E_y,E_z)$ and  $\ep_1(\omega)=\ep_r(1-\Omega_p^2/\omega^2)$ is the dielectric function 
of the medium within the Drude model~\cite{MarderBook,Kittelbook, GrossoBook} of free electron gas with $\ep_r$ as the relative background static dielectric constant of the TDM.
To be able to compare our results against Hofmann and coworkers~\cite{Hofmann2016}, we take $\ep_r=13$~\cite{SushkovEuIrO}.
Here $\Omega_p=4e^2\mu^2 /3\pi \hbar^3c$~\cite{MarderBook,GrossoBook} (in the CGS) is the plasmon frequency and the chemical potential $\mu$ fixes the electron density. 
From now on for convenience we write $\ep_1$ in place of $\ep_1(\omega)$. 
The boundary condition discussed under Eq.~\eqref{electric-t-c.eqn} requires that $E_y=0$ in both
sides of the surface. Therefore in the space of $(E_x,E_z)$, only corner elements of the above
matrix contribute. Requiring a non-trivial solution for $(E_x,E_z)$ by setting the 
pertinent determinant to zero gives
\be
\gamma_{1}^2=(q^2\lambda^2-\omega^2 c^{-2} \ep_1)/\lambda^2.
\label{decay-xx0.eqn}
\ee
For $\gamma_0$ we have
\be
   \gamma_0^2=q^2-\omega^2/c^{2},
   \label{gam0.eqn}
\ee
which expresses the propagation of a wave in $xy$ plane with wave vector $\bs q$ and its decay along $z$ with
a decay constant $\gamma_0$ into the vacuum. 
Finally the boundary conditions give another relation between $\gamma_0$ and $\gamma_1$:
\bearr
&&q^2\lambda^2+\ep_1(\gamma_1 \gamma_0-\omega^2 c^{-2})=0.
\label{dis-xx0.eqn}
\eearr

Substitution of $\gamma_1$ and $\gamma_0$ from Eqs.~\eqref{decay-xx0.eqn} and~\eqref{gam0.eqn} into Eq.~\eqref{dis-xx0.eqn} gives the
dispersion of SPP in tilted Dirac matter. 
In Fig.~\ref{fig2-xx0.fig} we have plotted the above dispersion relation for $\zeta=0$ (black solid line) and 
$\zeta=0.9$ (blue and green solid lines). As indicated in the legend, black solid curve corresponds to relative
dielectric constant $\ep_r=1$ while blue and green curves correspond to $\ep_r=1$ and $13$, respectively. 
The dashed line corresponding to every color represents the dispersion of electromagnetic and surface plasmons (flat part)
before they are coupled. All curves asymptote to their corresponding dashed curves. 
In general for an non-tilted ($\zeta=0$) system with relative dielectric constant $\ep_r$, the 
asymptotic value determining the surface plasmon frequencies is given by $\Omega_p/\sqrt{1+\ep_r^{-1}}$~\cite{GrossoBook}
which for $\ep_r=1$ reduces to the well celebrated relation due to Ritchie, $\Omega_s=\Omega_p/\sqrt 2$~\cite{Ritchie1957,GrossoBook}. 
Within this formula, the role of larger $\ep_r$ is to blue shift the $\Omega_s$ to bring it closer to $\Omega_p$. 
As can be seen in Fig.~\ref{fig2-xx0.fig}, by increasing $\zeta$ a similar effect can be achieved. 

\begin{figure}[t]
    \centerline{\includegraphics[width = .4\textwidth] {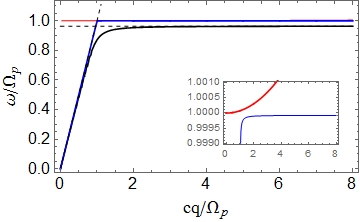}}
    \caption{(Color online) Same as Fig.~\ref{fig2-xx0.fig} but with $\zeta=0.999$. The inset shows that the
    bulk plasmons (red) continues to disperse weakly, but the SPP (blue) will be non-dispersive, corresponding to vanishing group velocity.}
    \label{fig3-xx0.fig}
\end{figure}

By comparison of black ($\zeta=0$) and blue ($\zeta=0.9$) curves, it can be clearly seen that
in the presence of tilt, the SPP dispersion asymptotes to energies larger than the corresponding upright Dirac semimetals ($\zeta=0$).
This blue shift of the surface plasmon frequency appears for arbitrary value of tilt parameter $\zeta$
and becomes stronger by approaching to the "horizon" value~\cite{Volovik2016,TohidBorophene,JafariEB} $\zeta=1$. 
As can be seen in Fig.~\ref{fig3-xx0.fig} in the $\zeta\to 1$ limit one has $\Omega_s\to\Omega_p$. This can be analytically understood as follows:
From the $q\to\infty$ limit of Eq.~\eqref{dis-xx0.eqn} we find that
\be
   \Omega_s=\Omega_p/\sqrt{1+(1-\zeta^2)/\ep_r}.
   \label{ExtendedRitchie.eqn}
\ee
This equation nicely generalizes the textbook result of Ritchie~\cite{GrossoBook} to include tilt of the environment. 
Eq.~\eqref{ExtendedRitchie.eqn} easily explains why in the horizon limit, $\zeta\to 1$, irrespective of the value
of the background dielectric constant $\ep_r$, always the surface plasmon frequency $\Omega_s$ approaches the 
bulk plasmon frequency $\Omega_p$. In addition to that, Eq.~\eqref{ExtendedRitchie.eqn} predicts that for
over-tilted WSMs with $\zeta>1$ the surface plasmon frequency can become even larger than the bulk plasmon 
frequency. Indeed for $1<\zeta<\sqrt{1+\ep_r}$, one has $\Omega_s>\Omega_p$. By approaching the limit $\zeta\to \sqrt{1+\ep_r}$
from left, the $\Omega_s$ diverges. Beyond this value, the SPP frequencies will become imaginary, and hence the
SPP modes will be unstable. These effects have not analog in non-tilted systems. 

Let us elaborate on another feature of Fig.~\ref{fig3-xx0.fig}. As can be seen, 
the smooth crossover from the light-like dispersion in the retarded limit ($q\to 0$) and the non-dispersive excitations
in the non-retarded limit ($q\to\infty$) becomes sharper as $\zeta$ approaches $1$. 
To show this more clearly in Fig.~\ref{fig3-xx0.fig} we have plotted SPP dispersion for $\zeta=0.999$. 
As a result of strong tilt, SPP (the blue line) and bulk plasmon (red line) dispersion come closer (see the inset).
The SPP becomes dispersionless and is pinned to $\Omega_s$ value  which from Eq.~\eqref{ExtendedRitchie.eqn} is 
equal to $\Omega_p$. That is why the red curve ($\Omega_p$) and blue curve ($\Omega_s$) join each other at the kink wave-vector. 
As can be seen in the inset, the bulk plasmons still have some dispersion, while the SPP modes become
dispersionless and therefore their group velocity becomes zero. 
This is striking feature, because it simply means that the SPP modes will not be able to propagate 
if their wave vector is beyond certain $q$ value corresponding to the kink in the dispersion of SPPs in this figure
which in the $\zeta=1$ limit is given by $cq_{\rm kink}/\Omega_p=1$.
This can be thought of as tilt induced filtering of the SPP modes with $q>q_{\rm kink}$, and the
inability of SPP modes to propagate for large wave vectors can be regarded as a plasmonic manifestation
of the "horizon" set by $\zeta=1$. 

Now let us assume that the tilt is in perpendicular to the surface and is given by $\bs\zeta=\zeta(0,0,1)$.  
In this case, despite that the representation of the matrix is different from Eq.~\eqref{tensor-xx.eqn}, but 
the resulting equations, namely Eq.~\eqref{decay-xx0.eqn},~\eqref{gam0.eqn} and~\eqref{dis-xx0.eqn} will be identical to 
the case with $\bs q||\bs\zeta$. 

The only remaining case is to consider $\bs q$ and $\bs\zeta$ both in the $xy$ interface, but $\bs q\perp \bs\zeta$. 
Strikingly the resulting dispersion relation is identical to the case where $\zeta=0$. Therefore in this case, there will be
no effect arising from the tilt in the SPP dispersion of TDM. 

\begin{figure*}[t]
   \includegraphics[width = \textwidth] {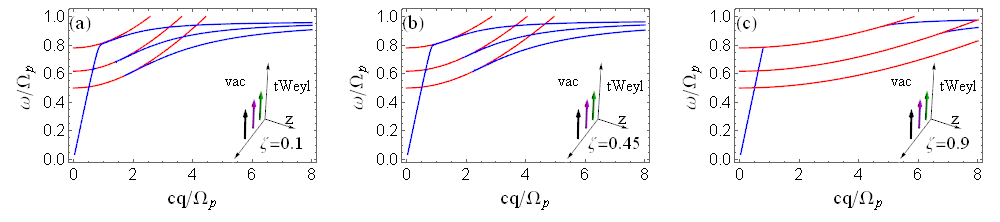}
   \caption{(Color online) The SPP dispersion on the TWM. The TWM occupies the $z>0$ region and the $xyz$ coordinate system is 
   right-handed. Arrows in the inset indicate momentum (green), $\bs b$ vector (purple) and tilt vector $\bs\zeta$ (black).
   In each panel curves from top to bottom correspond to $\omega_b=0.5,1,1.5$. Blue (red) curves correspond to SPP (bulk plasmon) 
   dispersion. Panels (a), (b) and (c) correspond to 
   tilt values $\zeta=0.1,0.45,0.9$, respectively.
   Since $\bs b$ is in the interface, the interface supports a Fermi arc whose length is set by $|\bs b|$. 
   }
   \label{fig4-xxx.fig}
\end{figure*}

\section{Tilted Weyl matter}{\label{sec4}}
Having discussed TDM with $b^\mu=0$, we are now ready to discuss the case $b^\mu\ne 0$ corresponding to TWM. 
This amounts to inclusion of the axion term in the Lagrangian, via a nonzero coupling constant $g_a$. 
In this section we include the axion term in the form of $b^\mu=(0,\bs b)$ where the temporal component is taken $b_0=0$ which
corresponds to Weyl semimetals with broken time-reversal symmetry.
We investigate role of both tilt and axion field on the SPP dispersion of the resulting tilted Weyl semimetals. 
As pointed out earlier, the effect of axion filed alone (without tilt) in the SPP of Weyl semimetals has been investigated 
by other~\cite{Hofmann2016,Polini2015,Tamaya2019}. Having fixed the direction $\hat z$ normal to the surface of TWM, in our case we will be
dealing with three vectors, namely, momentum $\bs q$,  axion field parameter $\bs b$ and tilt parameter $\bs\zeta$.
The situation where $\bs b$ is in the interface corresponds to a situation with Fermi arcs~\cite{Faraei2018Green,Faraei2019SCArc}.
The vector $\bs b$ can be chosen to lie in the interface plane or perpendicular to it by cutting the appropriate facet 
of the WSM and the existence of Fermi arc can be characterized with photo emission experiments~\cite{Xu2015,Xu2016,Deng2016,Yun2016}.
The propagation direction $\bs q$ can be chosen to be along the Fermi arc or perpendicular to it. The only thing which is not
at our control in three dimensional Weyl systems is the direction of the tilt $\bs\zeta$. 
So in the following for a given relative orientation of $\bs q$ and $\bs b$, we explore possible orientations of the tilt $\bs\zeta$. 

\subsection{Propagation along Fermi arc: $\hat{\mathbf q}||{\mathbf b}$ }
\underline{ (i) $\bs\zeta||\hat x$:}
Let us start with the parallel momentum and $\bs b$ (Fermi arc) and take $\bsq=q\hat x$ and $\bs b=b \hat x$ on the surface of TWM.
Then we investigate role of tilt magnitude and its direction. The wave vector relation in Eq.~\eqref{Mtensor.eqn} can be written 
in the compact matrix form
\be
\bs T\bs E=\bs 0,
\ee
where the precise form of $\bs T$ tensor depends on the orientation of $\bs\zeta$. 
For $\bs\zeta=\zeta \hat x$ (parallel to $\bsq$ and $\bs b$) it has the following representation
\be
 \!\!\!\begin{bmatrix}
	 -\gamma_1^2\lambda^2-\omega^2c^{-2} \ep_1	&	 0	&	-iq\gamma_1\lambda^2 \\
  	 0	&	q^2\lambda^2-\gamma_1^2-\omega^2c^{-2} \ep_1	&	i \omega^2c^{-2} \ep_2\\
     	-iq\gamma_1\lambda^2	&	-i \omega^2c^{-2} \ep_2	&	q^2\lambda^2- \omega^2c^{-2} \ep_1\\
\end{bmatrix}
\label{tensor-xxx.eqn}
\ee
where $\ep_2=\ep_r\omega_b/\omega=8\pi  b g_a c/\omega$ which also defines the frequency $\omega_b$ proportional to the $b=|\bs b|$. 
Nonzero $b$ and hence $\omega_b$ encodes the chiral anomaly~\cite{Kharzeev2014,Fujikawa1979,Peskin,ZeeQFTBook}
and is proportional to the length of Fermi arc in the simplest possible model~\cite{Faraei2018Green,Witten}.
It can be seen that the role of $\omega_b$ (Fermi arc) is to induce off-diagonal terms proportional to $\ep_2$ in the tensor $\bs T$ above.
This is how chiral anomaly is reflected in SPP spectrum of WSMs. In our case further information about $\zeta$ is encoded into
the related parameter $\lambda^2=1-\zeta^2$ which modifies all diagonal elements and the off-diagonal elements in the upper right
and lower left corners of the matrix. The determinant of above tensor has two nontrivial solution for decay constant 
$\gamma_{1,+}$ and $\gamma_{1,-}$ which are given by,
\begin{widetext}
\be
2 \ep_1\lambda^2 \gamma_{1,\pm}^2=q^2 \ep_1\lambda^2 \lambda^{\prime2} -
\omega^2c^{-2} (\ep_1^2 \lambda^{\prime2} -\ep_2^2 \lambda^2 )
\pm\sqrt{
q^4 \ep_1^2 \zeta^4 \lambda^4 + 
\omega^4c^{-4} (\ep_2^2 + (\ep_1^2 - \ep_2^2) \zeta^2)^2 
-2q^2 \omega^2c^{-2} \ep_1 \lambda^2 (\ep_1^2 \zeta^4 -
\ep_2^2\lambda^2 \lambda^{\prime2})},
\ee
where $\lambda^{\prime2}=2-\zeta^2$. Imposing the boundary condition and defining $\gamma_{1,s}=\gamma_{1,+}+\gamma_{1,-}$ and 
$\gamma_{1,g}=\sqrt{\gamma_{1,+}\gamma_{1,-}}$ gives rise to the following SPP dispersion relation,
\be
\gamma_{1,g}^2\lambda^4 (\gamma_0 + \gamma_{1,s}) + \gamma_0\ep_1\lambda^2(\gamma_{1,s}^2 -  \gamma_{1,g}^2)
+ q^2 \lambda^2 (\gamma_0\lambda^2- \gamma_0\ep_1+\ep_1  \gamma_{1,s}) 
+ \omega^2c^{-2}\ep_1 (\gamma_0 (\ep_1 - \lambda^2) - \lambda^2\gamma_{1,s})=0.
\label{disp-xxx.eqn}
\ee
\end{widetext}
In the case of WSMs considered in Ref.~\cite{Hofmann2016} corresponding to $\zeta=0$ and consequently $\lambda^2=1$ and $\lambda^{\prime2}=2$, 
this relation reduces to the corresponding equation of the above reference\footnote{Note that in there are some typos in the equations of the above reference}. 
\begin{figure*}[t]
   \includegraphics[width = \textwidth] {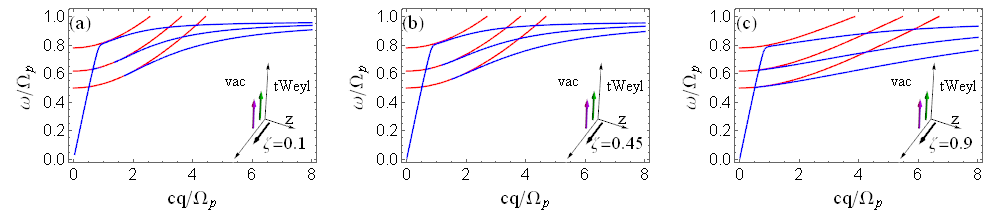}
   \caption{(Color online) Same as Fig.~\ref{fig4-xxx.fig}. As indicated in the legend,  the tilt $\bs\zeta$ is along $y$ axis. 
   The surface supports Fermi arc states. }
   \label{fig5-xxy.fig}
\end{figure*}
The SPP dispersion for different tilt parameter are represented in panels (a)-(c) of Fig.~\ref{fig4-xxx.fig}. 
In all panels, the SPP mode in the limit $cq/\om_p\ll1$ (fully retarded limit) shows the light like behavior and in opposite 
regime $cq/\om_p\gg1$ (non retarded limit) it approaches a constant value which is appropriate generalization of Richie constant~\cite{Ritchie1957}
as given by our formula~\eqref{ExtendedRitchie.eqn}. 
In each panel three curves from top to bottom correspond to $\omega_b=0.5,1,1.5$. The arrows in the inset represent the momentum (green), 
$\bs b$ vector encoding the Fermi arc (purple) and tilt vector $\bs\zeta$ (black). In this figure they are all along $x$ axis. 
Even in the absence of tilt, namely for $\zeta=0$, the right branch of the blue curves ceases to exist below a certain momentum.
This is a point where at least one of the decay constant $\gamma_{1,\pm}$ becomes imaginary, and therefore instead of SPP modes localized
on the surface one obtains bulk plasmons (red). That is why in all panels right at the point where SPP ceases to exist, 
the bulk plasmon (red branch) takes over. For large enough axion parameter $\omega_b$, there will be a gap in wave vectors
where SPP modes are not supported anymore. By increasing $\omega_b$, this SPP-forbidden region extends to larger wave vectors~\cite{Hofmann2016}.
At $\zeta=0$, this SPP-forbidden region can be attributed to the Fermi arc.
This picture holds qualitatively for small $\zeta$ as in panel (a) corresponding to $\zeta=0.1$. By further increasing $\zeta$
in panels (b) and (c), it can be seen that the SPP-forbidden region extends further. In panel (c) with $\zeta=0.9$, already for axion
parameter $\omega_b=0.5$ the SPP-forbidden region exists. 

Therefore as long as the formation of SPP-forbidden region is concerned, when $\bs b$ and $\bs \zeta$ are parallel and both lie
in the interface plane, they reinforce each other information of SPP-forbidden region and giving more room for bulk plasmons. 
But this is not a simple additive effect, as $\bs b$ appears in the off-diagonal element $T_{12}$ of tensor $\bs T$ of
Eq.~\eqref{tensor-xxx.eqn}, while $\bs \zeta$ affects the $T_{13}$ off-diagonal element and all diagonal elements. In fact, with respect to the
energy of SPP modes, these two parameters have quite opposite effects: Increasing the $\omega_b$ lowers the energy of SPP,
while increasing $\zeta$ increases the energy of SPP branch. This effect is generic and can be seen in all three panels of Fig.~\ref{fig4-xxx.fig}. 

\underline{ (ii) $\bs\zeta||\hat y$:}
In the second part of this subsection dealing with $\bs q||\bs\zeta$. 
We now consider a situation where $\bs\zeta\perp \bs q||\bs b$ but the tilt vector $\bs\zeta$ is still in the interface plane. 
Having fixed $\bs q$ and $\bs b$ to be along $x$ direction, we consider the direction of tilt vector is in the $y$ direction. 
In this case, the $\bs T$ tensor is
\be
\!\!\! \begin{bmatrix}
	 -\gamma_1^2-\omega^2c^{-2} \ep_1	&	 0	&	-iq\gamma_1 \\
  	 0	&	(q^2-\gamma_1^2)\lambda^2-\omega^2c^{-2} \ep_1	&	i \omega^2c^{-2} \ep_2\\
     	-iq\gamma_1	&	-i \omega^2c^{-2} \ep_2	&	q^2- \omega^2c^{-2} \ep_1\\
\end{bmatrix}
\label{tensor-xxy.eqn}
\ee
Consequently decay constant $\gamma_{1,\pm}$ are given by,
\begin{widetext}
\be
2 \ep_1\lambda^2 \gamma_{1,\pm}^2=
\pm \sqrt{
\omega^4c^{-4}  (\ep_2^2 - \ep_1^2 \zeta^2)^2 
+4 \omega^2c^{-2} q^2 \lambda^2 \ep_1 \ep_2^2  }
+2q^2 \ep_1\lambda^2+
\omega^2c^{-2} (\ep_2^2 -\ep_1^2 \lambda^{\prime2} ),
\ee
which upon imposing the boundary condition give the SPP dispersion as
\be
\gamma_{1,g}^2  (\gamma_0 + \gamma_{1,s} ) + \gamma_0\ep_1(\gamma_{1,s}^2 - \gamma_{1,g}^2)
+ q^2  (\gamma_0(1- \ep_1)+\ep_1 \gamma_{1,s} ) 
+ \omega^2c^{-2}\ep_1 (\gamma_0 (\ep_1 - 1) - \gamma_{1,s})=0.
\ee
\end{widetext}

The above SPP dispersion has been shown in Fig.~\ref{fig5-xxy.fig}.
In this figure color notation is same as Fig.~\ref{fig4-xxx.fig} i.e. the blue and red color stand for SPP and bulk plasmon dispersion respectively. 
The color code for $(\bs q,\bs b,\bs\zeta)$ vectors are also same as Fig.~\ref{fig4-xxx.fig}. 
As can be seen in this figure, the only difference in the orientations of the above vectors 
with respect to Fig.~\ref{fig4-xxx.fig} is that the $\bs\zeta$ vector is now aligned in $y$ direction. 
The values of $\omega_b=0.5,1,1.5$ again increase from top to bottom and the tilt parameter from panel (a) to (c) is given by $\zeta=0.1,0.45,0.9$, respectively.

\begin{figure}[t]
   \centerline{\includegraphics[width = .4\textwidth] {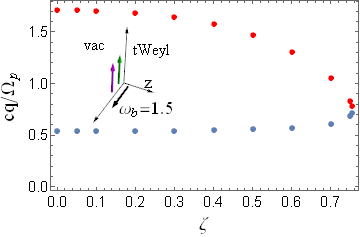}}
   \caption{(Color online) Trend of upper (red) and lower (blue) bounds of the SPP-forbidden region
   for a fixed $\omega_b=1.5$. 
   }
   \label{fig6-q-xxy.fig}
\end{figure}

For small tilt parameters in panels (a) and (b), still a large enough $\omega_b$ opens up a SPP-forbidden gap which
expands by increasing $\omega_b$. Focusing on the third curve from top in each of the panels (a) to (c), 
it can be discerned that increasing tilt parameter shrinks the SPP-forbidden window and in the limit of strong tilt, 
the SPP-forbidden window will totally disappear. 
Therefore a $\zeta$ transverse to Fermi arc that lies in the arc plane "repairs" the SPP-forbidden region
caused by strong rough $\omega_b$:
Whenever $\zeta$ is small, like in panel (a), the vector $\bs b$ gives rise to a SPP-forbidden window 
and whenever the tilting is strong, the $\bs\zeta$ will become dominant the forbidden window will be smaller. This trend has been plotted
in Fig.~\ref{fig6-q-xxy.fig} where by increasing $\zeta$, the two ends of the SPP-forbidden region denoted by blue and red
circles come closer to each other, and beyond a certain $\zeta$, the SPP region is closed. 

\begin{figure}[b]
   \centerline{\includegraphics[width = .4\textwidth] {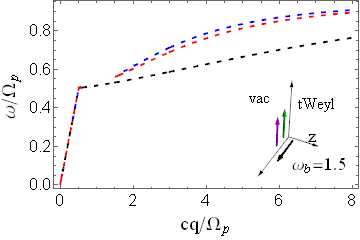}}
   \caption{(Color online) The SPP dispersion in the tilted Weyl for fixed $\omega_b=1.5$ and $\zeta=0$ (blue), $0.45$ (red)
   and $0.9$ (black). }
   \label{fig7-xxy-fwb.fig}
\end{figure}

Furthermore, increasing tilt parameter brings the non-retarded portion of the SPP dispersion curves to lower energies. 
To emphasize this trend, in Fig.~\ref{fig7-xxy-fwb.fig} we have plotted the SPP dispersions corresponding to a fixed Fermi arc parameter
$\omega_b=1.5$, for $\zeta=0,0.45,0.9$. As can be seen the retarded branch for all three values of $\zeta$ shown here coincide, while the non-retarded 
part is red-shifted to lower energies and the SPP-forbidden region shrinks by increasing $\zeta$. 
Furthermore for the large value of $\zeta=0.9$ where the SPP-forbidden region has disappeared, a kink in the 
dispersion develops. This feature is similar to the case of TDMs. But the difference is that in the case of 
TDMs, the group velocity to the right of $q_{\rm kink}$ vanishes by approaching $\zeta\to 1$, while in the
case of TWM they continue to {\em coherently} disperse at a {\em constant velocity}. 
This feature maybe desirable condition if one is interested in formation of a coherent state.
This configuration can be promising for SPP based lasers~\cite{Berini2011}.

\begin{figure*}[t]
   \includegraphics[width = \textwidth] {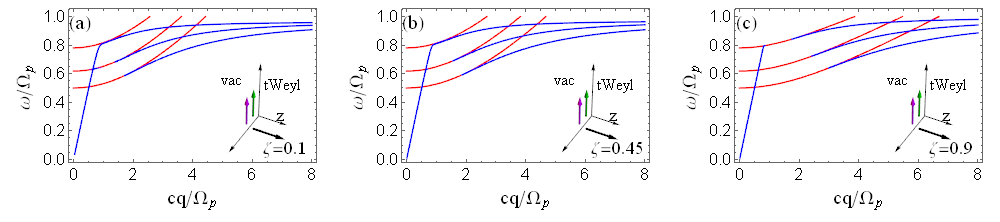}
   \caption{(Color online) Same as Fig.~\ref{fig4-xxx.fig}. As indicated in the legend,  the tilt $\bs\zeta$ is along $z$ axis. }
   \label{fig8-xxz.fig}
\end{figure*}

\underline{ (iii) $\bs\zeta||\hat z$:}
Having fixed $\bs q$ and $\bs b$, the last remaining possibility for $\bs\zeta$ is to point along $z$ direction. 
In this case the tensor $\bs T$ will be correspondingly modified. The resulting dispersion relation turns out to be
identical to the one given by Eq.~\eqref{disp-xxx.eqn}, but the decay constants will be

\begin{widetext}
\bearr
&&2 \ep_1 \lambda^2\gamma_{1,\pm}^2=\omega^2c^{-2} (\ep_2^2-2 \ep_1^2  ) +
  q^2 \ep_1 \lambda^{\prime2} \pm \sqrt{
  \omega^4c^{-4} \ep_2^4 + q^4 \ep_1^2 \zeta^4 +
   2 q^2 \omega^2 c^{-2}\ep_1 \ep_2^2 \lambda^{\prime2}}.
\eearr
\end{widetext}

The resulting SPP and bulk plasmon dispersion is plotted in Fig.~\ref{fig8-xxz.fig}. The trend in terms of the energy scale of SPP curves,
as well as the SPP-forbidden region is similar to Fig.~\ref{fig4-xxx.fig}. The difference is that in Fig.~\ref{fig4-xxx.fig}, the tilt parameter
$\zeta$ gives rise to a much wider SPP-forbidden region than the present figure. These two figures are in sharp contrast to $\bs\zeta=\zeta \hat y$ case
in Fig.~\ref{fig5-xxy.fig} where the tilt parameter restores the SPP region in entire range of momenta. 

To conclude this subsection on propagation along Fermi arc ($\bs q||\bs b||x$), in cases (i) and (iii) with $\bs\zeta||x$ and $\bs\zeta||z$, respectively, 
the tilt enhances the effect of $\omega_b$ in forming the SPP-forbidden region. This effect is stronger in the former case. 
In case (ii) with $\bs\zeta||y$ the $\bs\zeta$ nullifies the effect of $\omega_b$ and eliminates the SPP-forbidden window.

\subsection{Propagation transverse to Fermi arc: $\hat{\mathbf q}\perp {\mathbf b}$}
The vector $\bs q$ is always in the $xy$ plane of the interface. The parameter $\bs b$ of the
axion field can be still in the $xy$ plane, but perpendicular to $\bs q$. 
In this case still the surface states are Fermi arc states, but the propagation
direction is transverse to the Fermi arc direction.
Choosing $\hat{ \bs q}||\hat y$ and
$\hat{\bs b}||\hat x$, there will be three possibilities for the vector $\bs\zeta$ of the tilt. 

\underline{(i) $\bs\zeta||\hat x$}: 
We start by tilt direction along the $x$ axis which means that $\bs\zeta ||\bs b$. In this case the $\bs T$ tensor is given by
\be
 \begin{bmatrix}
	\lambda^2(q^2 -\gamma_1^2)	&	 0	& 0 \\
  	 0	&	-\gamma_1^2	&		-iq\gamma_1+i \frac{\omega^2}{c^{2}} \ep_2\\
     0	&		-iq\gamma_1-i \frac{\omega^2}{c^{2}} \ep_2	&q^2\\
\end{bmatrix}
-\frac{\omega^2}{c^{2}} \ep_1 \bs1.
\label{tensor-yxx.eqn}
\ee
As it can be seen from the above tensor, the only matrix element which is dependent on the tilt parameter is $T_{11}$.
But this block of the matrix is the trivial root for determinant. Hence in this case, 
the decay constant (coming from the lower $2\times 2$ block) will be independed of the tilting:
\be
 \ep_1\gamma_{1}^2=
q^2 \ep_1 +\omega^2c^{-2}(\ep_2^2-\ep_1^2).
\label{decay-yxx.eqn}
\ee
Moreover the boundary condition does not show any tilt dependence, and consequently the SPP dispersion as can be seen
in the following equation, does not recessives any correction from tilt parameter:
\be
q(q+ \ep_2 \gamma_0) +\ep_1( \gamma_1 \gamma_0 - \omega^2c^{-2})=0.
\label{disp-yxx.eqn}
\ee
The above decay constants and dispersion relation coincides with the corresponding equations for
non-tilted WSMs having $\zeta=0$~\cite{Hofmann2016}. Also the non-reciprocity effect  arising from the odd power of $q$ in 
Eq.~\eqref{disp-yxx.eqn} remains identical to the WSMs. 

\underline{(ii) $\bs\zeta||\hat y$ and (iii) $\bs\zeta|| \hat z$}: These remaining cases have the following representations for their $\bs T$ tensor
\be
\!\!\!\ \begin{bmatrix}
q^2	\lambda^2 -\gamma_1^2&	 0	& 0 \\
  	 0	&	-\gamma_1^2	\lambda^2	&		-iq\gamma_1	\lambda^2+i \frac{\omega^2}{c^{2}} \ep_2\\
     0	&		-iq\gamma_1	\lambda^2-i \frac{\omega^2}{c^{2}} \ep_2	&q^2	\lambda^2\\
\end{bmatrix}
-\frac{\omega^2}{c^{2}} \ep_1 \bs1,
\label{tensor-yxy.eqn}
\ee
and
\be
\!\!\! \begin{bmatrix}
q^2	-\gamma_1^2 \lambda^2&	 0	& 0 \\
  	 0	&	-\gamma_1^2	\lambda^2	&		-iq\gamma_1	\lambda^2+i \frac{\omega^2}{c^{2}} \ep_2\\
     0	&		-iq\gamma_1	\lambda^2-i \frac{\omega^2}{c^{2}} \ep_2	&q^2	\lambda^2\\
\end{bmatrix}
-\frac{\omega^2}{c^{2}} \ep_1 \bs1,
\label{tensor-yxz.eqn}
\ee
respectively. These tensors only differ in their trivial block (element $11$). The non-trivial block responsible for
the SPP dispersions in these two cases are identical. 
Hence both decay constant and boundary condition for the above two cases will be the same 
and are given by, 
\bearr
&& \lambda^2\ep_1\gamma_1^2= q^2 \ep_1 \lambda^2+\omega^2 c^{-2}(\ep_2^2 - \ep_1^2),\\
&&q( q+\ep_2 \gamma_0) +\ep_1( \gamma_1 \gamma_0 - \omega^2c^{-2}-\zeta^2q^2)=0.
\label{dispyxz.eqn}
\eearr

\begin{figure}[t]
   \centerline{\includegraphics[width = .4\textwidth] {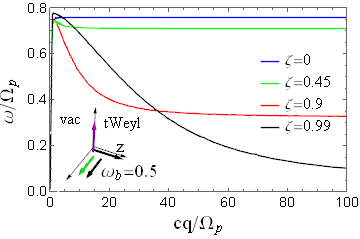}}
   \caption{(Color online) The SPP dispersion in the TWM for $\bs q||y$, $\bs b||x$. 
   This plot will be the same for both $\bs\zeta||z$ and $\bs\zeta||y$. 
   This figure is produced for $\omega_b=0.5$ and tilt parameters are $\zeta=0$ (blue), $\zeta=0.45$ (green), $\zeta=0.9$ (red) and
   $\zeta=0.99$ (black). 
   }
   \label{fig9-yxz.fig}
\end{figure}

In Fig.~\ref{fig9-yxz.fig} we have investigated the role of $\zeta$. The blue ($\zeta=0$), green  ($\zeta=0.45$), red  ($\zeta=0.9$) and black  ($\zeta=0.99$) 
curves represent the SPP dispersion in TWMs for $\omega_b=0.5$. This figure is the same for both $\bs \zeta|| y$ and $\bs \zeta || z$ cases. 
As can be seen, in these cases, the tilt has a very dramatic effect and causes a downturn in the SPP dispersion. Beyond the maximum, 
the branch which is continuously connected to non-retarded branch disperses downward in the energy and saturates at a $\zeta$ dependent
value which from the $q\to\infty$ limit turns out to be given by the following formula:
\be
\Omega_s=\frac{-\omega_b \ep_r + \sqrt{
 \omega_b^2 \ep_r^2 + 
  4 \Omega_p^2 \lambda^2 \ep_r (1 + \ep_r \lambda^2)}}{2 (1 + \ep_r \lambda^2)}. 
  \label{tiltOmegas.eqn}
\ee
\begin{figure}[b]
   \centerline{\includegraphics[width = .4\textwidth] {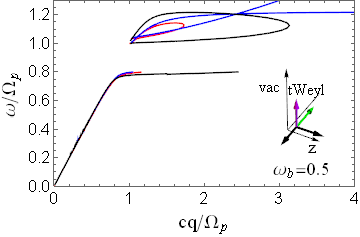}}
   \caption{(Color online) Nonreciprocity effect in TWMs: Note that the difference of the arrangements of $\bs q$, $\bs b$ and $\bs\zeta$ 
   in this figure and Fig.~\ref{fig9-yxz.fig}. is the reversal of the momentum (green) vector in the inset. 
   To be comparable with the results of Ref.~\cite{Hofmann2016}, this figure is produced for $\omega_b=0.5$.
   The values of $\zeta=0,0.45,0.9$ respectively correspond to blue, red and black curves.
   }
   \label{fig10-yxz-q.fig}
\end{figure}
As can be seen in the "horizon limit", $\zeta\to 1$ ($\lambda\to 0$), the above surface plasmon frequency 
approaches to zero and becomes a soft mode in the large $q$ limit. 
The softening of the surface plasmon frequency is a unique manifestation of the tilt, and has no analog in 
non-tilted systems. 
A gapless surface plasmon mode in the short wavelengths ($q\to\infty$) is quite unusual.
This qualifies SPP modes as appropriate quasi particles not only in the long-wavelength limit, but also in the
atomic length scales. This can be viewed as a unique effect of tunning $\zeta$ to its "horizon limit", $\zeta\to 1$.

Other notable features of the SPP dispersion in this case are: (1) At large momenta, the group velocity is negative.  
(2) Neither the axion parameter $\bs b$, nor $\bs \zeta$ in this case does produce an SPP forbidden region. 
So we have a whole branch of SPP excitations with unusual properties. (3) There is also non-reciprocity effect
due to odd powers of $q$ which is shown in Fig.~\ref{fig10-yxz-q.fig}. As can be seen in this figure, the low-momentum
branch of the SPP dispersion ceases slightly after the linear dispersion ends. The effect of tilt is to extend the 
point at which SPP dispersion ceases, to larger momenta. This is in contrast to Fig.~\ref{fig9-yxz.fig} where 
the dispersion continues upto unlimited positive $q$ values. 
At energies larger than the bulk $\Omega_p$, there appear
multiple roots. Note that in normal metals, the SPP eigenvalue equation requires
the dielectric function $\ep(\Omega_s)<0$. But in the present situation of TWMs,
the eigenvalue problems are more complicated, and there can be solutions even for
energy scales with $\omega>\Omega_p$. These types of solutions can be clearly seen 
in Fig.\ref{fig10-yxz-q.fig}. Therefore in this configurations, the range of SPP
energies can be further widened to energies above the bulk plasmon frequencies.

\begin{figure*}[t]
   \includegraphics[width = \textwidth] {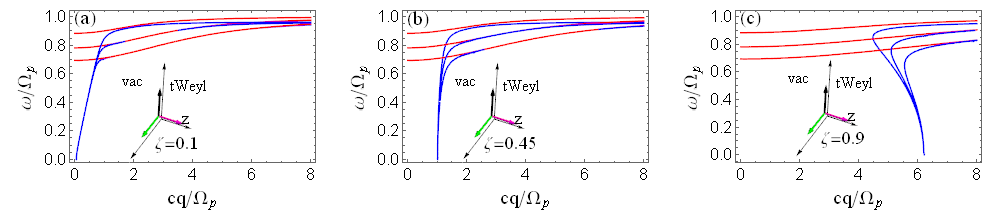}
   \caption{(Color online) Conventions are same as Fig.~\ref{fig4-xxx.fig}. In each panel curves from top to bottom correspond to 
   $\omega_b=0.25,0.5,0.75$. Panels (a), (b) and (c) correspond to tilt values $\zeta=0.1,0.45,0.9$, respectively. The relative
   directions of $(\bs q,\bs b,\bs\zeta)$ are indicated in the legend. }
   \label{fig11-yzx.fig}
\end{figure*}

\subsection{Interface without Fermi arc: $\hat{\mathbf q}\perp {\mathbf b}$}
The final situation we consider is that the vector $\bs b$ is still perpendicular to $\bs q$, but $\bs b$ is not 
in the plane of the interface anymore. So we assume $\bsq=q\hat y$ and $\bs b=b \hat z$. 
In this case the interface {\it does not support Fermi arcs}. 
For the tilt parameter $\bs\zeta$ there are again three options.

\underline{(i) $\bs\zeta||\hat x$:}
In this case, the decay constant and SPP dispersion relation will be determined from the following equations:
\begin{widetext}
\bearr
  &&\gamma^2_{1,\pm}=q^2- \frac{\omega  c^{-1}}{2\sqrt{\ep_1}\lambda^2}
  [\omega  c^{-1} \lambda^{\prime2}\ep_1^{3/2}  \pm \sqrt{
        \omega^2 c^{-2} \ep_1 (\ep_1^2 \zeta^4 +
           4 \ep_2^2\lambda^2)  - 4 q^2 \ep_2^2\lambda^2 }],\\
 &&q^2\lambda^2 [\gamma_{1,g}^2  + \gamma_0 \gamma_{1,s} +
    \ep_1 (\gamma_{1,s}^2 - \gamma_{1,g}^2)] 
     -  
 \omega^2c^{-2}\ep_1 [\gamma_{1,s} \gamma_0 + \gamma_{1,s}^2 + q^2 (1 - \ep_1)] 
 +\omega^2c^{-2}\ep_1 \zeta^2 [\gamma_{1,g}^2 + \gamma_0\gamma_{1,s}  +\ep_1 (\gamma_{1,s}^2 - \gamma_{1,g}^2) ]  \nn\\&&+
  q^4 \lambda^2 (1 - \ep_1)  +\ep_1  \lambda^2 \gamma_{1,g}^2 \gamma_0 \gamma_{1,s}=0.
\eearr
\end{widetext}

In this case where no Fermi arcs are supported, the role of tilt parameter is drastically different from previous cases supporting Fermi arcs.
As can be seen in Fig.~\ref{fig11-yzx.fig}, the most drastic effect of tilt $\zeta$ is to prevent formation of a solution at small $q$. 
This effect is enhanced by further increasing $\zeta$ from $0.1$ to $0.45$ in panel (b). This means that the role of tilt is to 
prevent propagation of SPP modes whose wavelength are larger than a certain $\lambda_{\rm min}$. An implicit equation for this quantity
can be constructed by setting $\omega=0$ in the above equations and agrees with numerical results presented in Fig.~\ref{fig11-yzx.fig}. 
Another unusual feature that can be noticed by comparing panels (a), (b) and (c) is that, upon increase in $\zeta$, not only $\lambda_{\rm min}$
is pushed to lower wavelengths, but also the phase velocity diverges and in panel (c) becomes a very large negative number. 
In addition to that, the absolute value of the phase velocity at larger $\zeta$ exceeds the light velocity. 
This is because at the larger $q$, the non-retarded nature encoded into the plasmons (via the non-retarded dielectric 
function) becomes dominant.
Such a tilt-induced surface plasmon polariton effect has no analog in non-tilted systems possessing standard Minkowski geometry. 
 
\begin{figure}[t]
\includegraphics[width =.5 \textwidth] {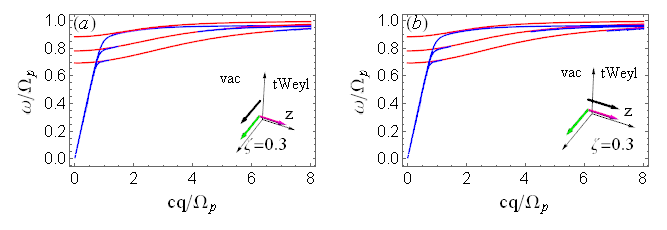}
    \caption{(Color online)  The arrows and $\omega_b$ conventions are same as Fig.~\ref{fig11-yzx.fig}. 
    As indicated in the legend,  the tilt $\bs\zeta$ in (a) is along $y$ axis 
    and in (b) is along $z$ axis. The tilt parameter in both graphs is $\zeta=0.3$. }
    \label{fig12-yz-yz.fig}
 \end{figure}

\underline{(ii) $\bs\zeta||y$:} For $\zeta$ along the momentum vector the following equations
express the decay constant and SPP dispersion relation:
\bearr
&&2 \lambda^2 \gamma_{1,\pm}^2=-
 \lambda^{\prime2} (\omega^2 c^{-2} \ep_1 - q^2 \lambda^2) \pm\\&&
 \sqrt{(\omega^2 c^{-2} \ep_1 - q^2 \lambda^2) (
   \omega^2 c^{-2} (\ep_1^2 \zeta^4 +
      4 \ep_2^2 \lambda^2)-  q^2 \ep_1 \zeta^4\lambda^2 )/\ep_1},\nn\\
&& q^2 \lambda^2 [\gamma_{1,g}^2 + \gamma_0 \gamma_{1,s}] +
 q^2   \ep_1 (\gamma_{1,s}^2 - \gamma_{1,g}^2) -
 \nn\\&&    
 \omega^2c^{-2}\ep_1 [\gamma_{1,s} \gamma_0 + \gamma_{1,s}^2+ 
    q^2 (\lambda^2 - \ep_1)] +
 \nn\\&&
  q^4 \lambda^4 - 
 q^4 \zeta^2 \lambda^2 \ep_1+\ep_1   \gamma_{1,g}^2\gamma_0 \gamma_{1,s}=0.
\eearr
The solutions of these equations is shown in panel (a) of Fig.~\ref{fig12-yz-yz.fig}. 
In this case the tilt does not gap out the low-$q$ part of the SPP dispersion. 
But still the typical behavior that the SPP-forbidden region expands by 
increasing $\omega_b$, as in Fig.~\ref{fig8-xxz.fig}, still holds here. 
Furthermore, for a fixed $\omega_b$, increasing $\zeta$ enlarges the SPP forbidden region (not shown here). 
  
 \underline{(iii) $\bs\zeta||\hat z$:}
Finally when $\bs\zeta$ is along $z$ direction, it will be parallel to the axion field parameter $\bs b$. 
In this case the decay constant and dispersion relation are given by,
 \bearr
&&2 \lambda^2 \gamma_{1,\pm}^2=q^2 \lambda^{\prime2}-
  2\omega^2 c^{-2} \ep_1 \\&& 
  \pm  \sqrt{(
  4 \omega^4 c^{-4} \ep_1\ep_2^2+q^4\ep_1 \zeta^4 -
      4 q^2 \omega^2 c^{-2} \ep_2^2 \lambda^2)/\ep_1}\nn\\
       \nn\\
 &&q^4 (\lambda^2- \ep_1)+ q^2 \lambda^4 (\gamma_{1,g}^2 + \gamma_0 \gamma_{1,s}) + \ep_1 \lambda^2  \gamma_{1,g}^2\gamma_0 \gamma_{1,s}\nn\\&&
  -\omega^2c^{-2}\ep_1 [\lambda^2\gamma_{1,s} (\gamma_0 + \gamma_{1,s}  )+ q^2 (\lambda^2 - \ep_1)] \nn\\&&
   +q^2    \lambda^2 \ep_1 (\gamma_{1,s}^2 - \gamma_{1,g}^2)    
  =0
 \eearr
  
For this case, the SPP dispersion has been shown in panel (b) of Fig.~\ref{fig12-yz-yz.fig}.
Again, the enhancement of the $\omega_b$ (from top to bottom) increases the SPP-forbidden area.
But in this case there is an important difference that in contrast to the behavior of Fig.~\ref{fig8-xxz.fig}
where increasing the tilt $\zeta$ expands the SPP-forbidden region. 
The SPP-forbidden region finally goes away for strong enough tilt parameter $\zeta$. 
The decreasing behavior of SPP-forbidden region is similar to Fig.~\ref{fig5-xxy.fig}. 
That is why in panel (b) of Fig.\ref{fig12-yz-yz.fig} we have
only presented the SPP dispersion for a fixed value of $\zeta=0.3$. 

\begin{figure}[t]
 \includegraphics[width =.4 \textwidth] {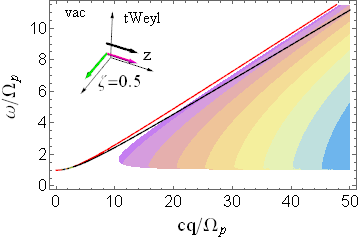}
     \caption{(Color online)  Conventions for arrow are as previous figures.  
     As indicated in the legend, the tilt $\bs\zeta$ (black arrow) and $\bs b$ (purple arrow) are both along $z$ axis. 
     In this figure $\zeta=0.5$ and $\omega_b=0.25$. 
     In white regions, imaginary part of $\gamma$'s are non-zero. In the shaded colored region, only real and positive $\gamma$ solutions are possible. 
     Red (black) curve is bulk plasmon (SPP mode). For large enough momenta, SPP mode with real and positive $\gamma$ emerges. 
     }
     \label{fig13-yzz+1.fig}
  \end{figure}

Strikingly, in this case an additional gaped SPP branch at high energies, i.e. $\omega/\Omega_p>1$ emerges. 
This has been shown in Fig.~\ref{fig13-yzz+1.fig}. The color shaded region in this figure corresponds to 
real and positive values of $\gamma$. The black (red) curve corresponds to SPP (bulk plasmon) dispersion. 
This curve is produced for $\omega_b=0.25$ and $\zeta=0.5$. 
As can be seen for large enough momenta, the SPP branch enters the region of real and positive $\gamma$ values
and therefore are stable SPP solutions. This high energy SPP branch {\em does not} exist for ordinary WSMs with $\zeta=0$. 
Even when $\zeta$ is small enough, this branch does not exist. Moreover, for very strong $\zeta$ as well, it does not
exist. Therefore this branch is a genuine effect of the tilt  parameter and emerges only at a restricted window of tilt values.

\begin{figure}[b]
\includegraphics[width =.4 \textwidth] {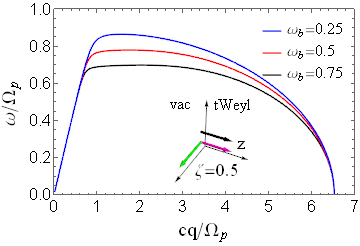}
    \caption{(Color online) Conventions for arrows are as before. 
    The tilt value is $\bs\zeta=0.9$ and values of $\omega_b$ as indicated in the legend
    are $0.25, 0.5$ and $0.75$ corresponding to blue, red and black curves. 
    }
    \label{fig14-yzz-gapless.fig}
\end{figure}

There is yet another aspect of the SPP dispersion in the present case. 
This happens when we fix the tilt at a large value of $\zeta=0.9$ as in Fig.~\ref{fig14-yzz-gapless.fig}.
In this case a quite unusual behavior emerges: The SPP curves, instead of asymptotically approaching $\Omega_s$
bend down and hit the zero energy at a finite $q_{\rm inst}$ which indicates and instability at
SPP excitations. It means that, beyond this wave vector, the SPP can not propagate. 
An implicit expression for $q_{\rm inst}$ can be obtained by setting $\omega=0$ in the SPP eigenvalue 
problem and discarding the $q=0$ solution. This turns out to be independent of $\omega_b$ and
depends on $\zeta$. Therefore $q_{\rm ins}=q_{\rm ins}(\zeta)$. Numerical solution of this
equation shows that the instability wave vector is a decreasing function of $\zeta$ and
saturates at $cq_{\rm ins}=\sqrt{\ep_r}\Omega_p=3.60555\Omega_p$ when $\zeta=1$ due to the following equation
\be
   q \Omega_p^2 \ep_r ( q^3 - \Omega_p^3 \sqrt{\ep_r^3} )=0.
\ee
This aspect of the present configuration is again like an SPP filter where SPP modes with 
smaller wavelengths are not sustained on the sample surface, and therefore can not
propagate through the sample.

\section{broken inversion symmetry}{\label{sec5}}
So far we have assumed that the temporal component of the $b^\mu$ is zero ($b_0=0$) and considered it as a vector in 
three dimension and investigated the role of tilt vector, $\bs\zeta$ in TWMs. Now the question arises, how temporal part of  $b^{\mu}=(b_0,\bs 0)$ 
in the TWMs affects the SPP modes. In this case, since the spatial part of $\bs b$ is absent, there will remain only two 
vectors, namely $\bs q$ and $\bs\zeta$ the relative orientation of which should be considered. 
In the case of upright Weyl cones studied in Ref.~\cite{Hofmann2016}, since $\bs\zeta$ is zero, the only 
remaining vector will be $\bs q$. But in that case from the isotropy of the theory in the $xy$ plane, it 
follows that the SPP mode does not depend on the direction of $\bs q$. In this case, the main effect of 
$b_0$ that enters the equations through the combination $\omega_{b0}^2=2b_0  e^2/\pi \hbar c \ep_r$, is to create a 
SPP-forbidden region where the bulk plasmons take over, and the possibility of hybridization between SPP modes and bulk plasmons arises. 
When we turn on $\bs\zeta$, the tilt vector $\bs\zeta$ can either be parallel or perpendicular to $\bs q$.
But since $\bs\zeta$ is not confined to the plane of interface, in the perpendicular $\bs\zeta$ can be either in plane,
or out of plane. Let us now consider them in the following.

\underline{(i) $\bs q || \bs\zeta$:} In the parallel case which both tilt and momentum wave vector are placed in the interface, the dispersion relation and decay constant are expressed as:
\bearr
\label{paralelb0.eqn}
&&
     \omega^4 c^{-4} \ep_1 (\gamma_0 + \gamma_{1,s})-\omega^2 c^{-2} (\gamma_0+\gamma_{1,s})(q^2\ep_1+\lambda^2\gamma_{1,g}^2)
     \nn\\&& - \omega^2 c^{-2} \gamma_0\gamma_{1,g}^2\ep_1+ q^2 \gamma_0 \lambda^2 (\gamma_0^2 +2 \gamma_{1,g}^2-\gamma_{1,s}^2) 
     \nn\\&& +  \gamma_0  \gamma_{1,g}^4 \lambda^2
     =0,
\eearr
\begin{widetext}
\bearr
     &&2 \lambda^2\gamma_{\pm}^2=q^2 \lambda^2\lambda^{\prime2}-\omega^4 c^{-4} \ep_2^2 - \omega^2 c^{-2}  \ep_1 \lambda^{\prime2} \nn\\
     &&\pm\sqrt{
  \omega^8c^{-8} \ep_2^4 +
   2 \omega^6 c^{-6}  \ep_1 \ep_2^2 \lambda^{\prime2} 
 -  2 q^2 \omega^2 c^{-2}  \ep_1 \zeta^4 \lambda^2 +
   q^4 \zeta^4 \lambda^4 + 
   \omega^4 c^{-4}  \zeta^2 (\ep_1^2 \zeta^2 +
      2 q^2 \ep_2^2 \lambda^2)}.
\eearr
\end{widetext}

\begin{figure}[t]
    \includegraphics[width = .5\textwidth] {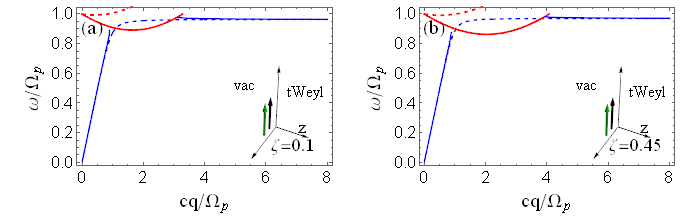}
    \caption{(Color online) The SPP dispersion in the tilted Weyl for $b^\mu=(b_0,{\bs 0})$. 
    The axion parameter appears through $\omega_{b0}$ which for solid (dashed) curves is $0.5$ ($0.25$).  
    The red (blue) curves as before correspond to bulk plasmon (SPP mode). 
    Panel (a) is for $\zeta=0.1$ while panel (b) is for $\zeta=0.45$. As before, the green and black arrows
    denote $\bs q$ and $\bs\zeta$. } 
    \label{fig15-b0-xx.fig}
 \end{figure}

In the above relations, we have assumed that $\ep_2=\ep_r \omega_{b0}^2/\omega^2$ and $\ep_1(\omega)=\ep_r(1-\Omega_p^2/\omega^2)$.
The qualitative behavior of SPP modes has been shown in Fig.~\ref{fig15-b0-xx.fig} with $\zeta=0.1$ in panel (a) and $\zeta=0.45$ in panel (b). Here the blue (red) curves correspond to the SPP (bulk plasmon) dispersion. In this figure the dashed (solid) curves correspond to $\omega_{b0}=0.25$ ($\omega_{b0}=0.5$). 
As before, the SPP-forbidden region is characterized by the fact that 
at least one of the decay constant $\gamma_1$ or $\gamma_2$ is a complex variable, so that the $z$ dependence
changes from decaying behavior (SPP) to propagating (bulk plasmon) behavior. 
By increasing $\omega_{b0}$, the SPP-forbidden region in which the bulk plasmon branch (red solid line) takes over, will be enlarged.
This aspect is quite similar to the dependence of SPP-forbidden region on the parameter $\omega_b$. 
Similar to the case of large $\omega_b$, a further increase in $\omega_{b0}$ -- beyond the values shown in this figure -- the SPP-forbidden region
will also expand from the left side and portions of the linear SPP dispersion in small $q$ will give way to the SPP-forbidden gap. 
Moving to the panel (b) of Fig.~\ref{fig15-b0-xx.fig} and comparing with panel (a) shows that
again as before, by increasing the tilt parameter $\zeta$, the SPP-forbidden region is enlarged. 
For example, for fixed value $\omega_{b0}=0.25$, beyond the tilt parameter $\zeta=0.796$, the SPP-forbidden region sets in. 
One should notice that for a constant $\omega_{b0}$, an increase in the tilt modifies the SPP-forbidden area by expanding it
from the higher-q side. But when the tilt becomes very strong, the SPP-forbidden region will expand from both lower-q and higher-q  sides. 

\underline{(ii) $\bsq \perp \bs \zeta$:}
In the remaining case, where the tilt direction is perpendicular to the momentum, the tilt parameter has two option: 
It can be either in or out of the interface plane. The dispersion relation of SPP for the situation where the tilt parameter 
is in the interface plane and perpendicular to the momentum (i.e. $\bs \zeta\perp \hat z$) is given by
\bearr
&&2 \lambda^2\gamma_{1,\pm}^2= 2 q^2\lambda^2-\omega^4 c^{-4} \ep_2^2 \nn \\&& +\omega^2 c^{-2} (-\ep_1 \lambda^{\prime2}
 \pm 
 \sqrt{
 \omega^4 c^{-4} \ep_2^4  +
  2 \omega^2 c^{-2} \lambda^{\prime2} \ep_1 \ep_2^2 + \ep_1^2 \zeta^4}),\nn\\
  \nn\\&&
q^4\gamma_{0} - q^2 \omega^2 c^{-2} [(1 + \ep_1)\gamma_{0} + 
    \ep_1 \gamma_{1,s}]  -  q^2 \gamma_0 (\gamma_{1,s}^2 -2 \gamma_{1,g}^2)
    \nn \\&&
 -\omega^2 c^{-2} \gamma_{1,g}^2  [(1 + \ep_1)\gamma_{0} +  \gamma_{1,s}] + \gamma_0 \gamma_{1,g}^4\nn\\&& + 
 \omega^4 c^{-4} \ep_1 (\gamma_0+ \gamma_{1,s} )=0.
\eearr
 When the tilt is normal to the interface and $\bs q$ (i.e. $\bs \zeta || \hat z$) the dispersion of SPP is same as Eq.~\eqref{paralelb0.eqn}, 
 but the decay constants satisfy the following equation:
 \bearr
 &&2 \lambda^4\gamma_{\pm}^2= q^2 \lambda^2 \lambda^{\prime2} -\omega^4 c^{-4}\ep_2^2 - 2 \omega^2 c^{-2} \ep_1 \lambda^2 \pm\\
 && \sqrt{
  \omega^8 c^{-8} \ep_2^4 +4  \lambda^2
    \omega^6 c^{-6}\ep_1 \ep_2^2 -2 \lambda^2
    q^2 \omega^4 c^{-4} \ep_2^2 \zeta^2 + \lambda^4
   q^4 \zeta^4 }.\nn
 \eearr 
\begin{figure}[t]
    \includegraphics[width = .4\textwidth] {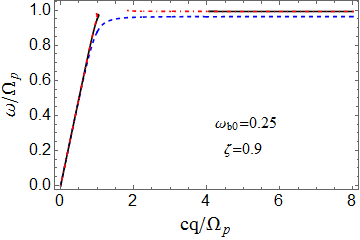}
    \caption{(Color online) The SPP dispersion in TWM in the presence of time component of
    axion filed parameter $b^\mu$ for a fixed values of $\omega_{b0}=0.25$ and $\zeta=0.9$. The momentum $\bs q$ is along $x$ direction.
    The tilt parameter for black, blue and red curves is along $x,y,z$ directions, respectively. }
    \label{fig16-all-b0.fig}
 \end{figure}
  
The qualitative  behavior of SPP dispersion will not differ much from Fig.~\ref{fig15-b0-xx.fig}. Therefore in Fig.~\ref{fig16-all-b0.fig},
we have chosen to represent the dispersion arising from all the three cases discussed above for fixed $\omega_{b0}=0.25$ and $\zeta=0.9$. 
In Fig.~\ref{fig16-all-b0.fig} $\bs q$ (green arrow in Fig.~\ref{fig15-b0-xx.fig}) is fixed along $x$ direction. The $\bs\zeta$ 
can be along $x$ (black), $y$ (blue) or $z$ (red) direction. The energy of surface plasmons denoted by the blue curve corresponding to $\bs\zeta$ along $y$ is
slightly less than the other two curves. By increasing tilt parameter, this energy difference will be enhanced. 
As can be seen in the blue curve there is no SPP-forbidden region. The black curve corresponding to $\bs\zeta||\bs q||\hat x$
has the largest SPP-region. Its behavior as a function of $\zeta$ is similar to the $\bs b\ne 0$ case. In blue curve (momentum and tilt perpendicular and in the interface), the situation is quite contrary
and the effect of tilt is not notable. Even a tilt as strong as $\zeta=0.9$ shown in this figure, is not 
able to open a SPP-forbidden gap. 
The trend with respect to the formation of SPP-forbidden region in the red curve corresponding to $\bs\zeta||z$ is something
intermediate between black and blue curves. 
So summarize, when $b^\mu=(b_0,\bs 0)$, the SPP-forbidden region for the parallel case is maximum and when $\bs\zeta=\zeta \hat z$ 
this region will be smaller. Finally when $\bs\zeta=\zeta \hat y$ for $\omega_{b0}=0.25$ case shown here, $\zeta=0.9$ has not been able to
create a SPP-forbidden region. 
 
 \section{Summary and Conclusion}{\label{sec6}}
In this work, we constructed a Maxwell electrodynamics theory for tilted Dirac/Weyl cone systems.
The modified maxwell equations are determined by two parameters: The axion field is parametrized by
$b^\mu=(b_0,\bs b)$ and the tilt is parameterized by $\bs\zeta=(\zeta_x,\zeta_y,\zeta_z)$. 
When $b^\mu=(0,\bs 0)$ the theory reduces to that of tilted Dirac matter, while for $b^\mu\ne 0$ we have
a Maxwell electrodynamics theory for the tilted Weyl matter. The essential way in which the tilt
parameter $\bs\zeta$ enters the electrodynamics of Weyl materials is through the metric of the spacetime, Eq.~\eqref{tmetric.eqn}. 
We found that the homogeneous Maxwell equations remain intact as in Eq.~\eqref{curlE.eqn} and~\eqref{divB.eqn}. 
But the inhomogeneous equations will be as given in Eq.~\eqref{rho.eqn} and~\eqref{j.eqn}. The striking 
modification in Eq.~\eqref{rho.eqn} is that the charge distribution $\rho$, not only affects the profile
of field $\bs E$, but also affects the profile of field $\bs B$ via the vector $\bs\zeta$. 
Similarly, in the modified Ampere's law, additional terms will appear which contain the temporal derivative 
of $\bs B$, and spatial derivatives of both $\bs E$ and $\bs B$ in a way summarized in Eq.~\eqref{j.eqn}. 

As an application of the above electrodynamics, we studied the surface plasmon polariton for a TDM or TWM
which interface with vacuum is the $xy$ plane. 
First, we considered a TDM corresponding to no axion field,
namely $b^\mu=0$. In this case the role of tilt parameter $\bs\zeta$ is to shift the surface plasmon frequency
according to Eq.~\eqref{ExtendedRitchie.eqn} which extends the Ritchie equation. Now the frequency $\Omega_s$ 
to which SPP modes asymptotically at large $q$ limit can approach the bulk plasmon frequency $\Omega_p$ as 
$\zeta\to 1$. In this limit, a sharp kink at wave vector $q_{\rm kink}$ in the SPP dispersion develops which
separates the retarded part with a coherent group velocity equal to $c$ from the localized (non-retarded) part
with zero group velocity. Furthermore, in the over-tilted situation $1<\zeta<\sqrt{1+\epsilon_r}$, the
surface plasmon frequency $\Omega_s$ exceeds the bulk plasmon frequency $\Omega_p$. This effect has 
no analog in other materials and is a unique feature of Maxwell theory for tilted cone systems.  
Furthermore at the value $\zeta=1$ corresponding to "horizon", the SPP modes cease to propagate for wave vectors
exceeding $q_{\rm kink}$ in Fig.~\ref{fig3-xx0.fig}. This can be viewed as a plasmonic manifestation of the black-hole 
horizon. 

In the case of TWM with either non-zero $\bs b$ or non-zero $b_0$, one needs to consider separate cases
depending on the relative orientation of three vectors $\bs b$, $\bs q$ and $\bs \zeta$.
The orientation of $\bs b$ can be controlled by deciding which facet of the WSM we want to cut. 
The propagation direction $\bs q$ is the direction in which we would like to propagate the surface
plasmon polariton. The direction of $\bs\zeta$ is fixed by the material at hand. 
Parameters $|\bs b|$ and $b_0$ enter the theory through frequencies
$\omega_b$ and $\omega_{b0}$. In the absence of tilt, the typical effect of both parameters is to create the SPP-forbidden region where
at least one of the decay constants acquires imaginary part whereby SPP modes cease to exist and propagating
bulk plasmons take over~\cite{Hofmann2016}. The generic effect of inclusion of $\zeta$ was to enlarge the 
SPP-forbidden region. 

Depending on whether the interface supports Fermi arcs or not, we considered three situations: \\
(A) Propagation along the Fermi arc, $\bs q||\bs b||x$ and $\bs \zeta||y$
shown in Fig.~\ref{fig5-xxy.fig} and Fig.~\ref{fig6-q-xxy.fig}. In this case increasing the tilt $\zeta$ closes the SPP-forbidden region. \\
(B) For SPP modes propagating transverse to Fermi arc, $\bs b||x$, $\bs q||y$, the results were shown in Fig.~\ref{fig9-yxz.fig}.
Then irrespective of whether $\bs\zeta||y$ or $\bs\zeta||z$, 
the SPP dispersion instead of saturation (with positive group velocity) at $\Omega_s$, bends downward first, and then subsequently saturates 
with negative group velocity to a heavily "tilt-renormalized" surface plasmon frequency given by Eq.~\eqref{tiltOmegas.eqn}. 
This new saturation velocity tends to zero when the tilt is tunned to $\zeta\to 1$. This strange behavior means that
at the "horizon value" of $\zeta=1$, soft modes emerge at large $q$. These are unlike the familiar soft modes that arise
in the long wavelength $q\to 0$ limit. The $q\to\infty$ soft mode at $\zeta=1$ would therefore, imply low-energy SPP
excitations at the atomic scales. This can be viewed as plasmonic manifestation of "horizon".  \\
(C) Surface does not support Fermi arcs, $\bs q \perp \bs b \perp \bs\zeta||z$ in Fig.~\ref{fig11-yzx.fig}.
In this situation, the tilt parameter filters the soft part of the SPP dispersion in the long wavelength side. 
Therefore the Fermi arcs play an important role in facilitating the propagation of long-wavelengths SPP modes. 
Other fascinating observation in this case is shown in Fig.~\ref{fig14-yzz-gapless.fig} where the SPP dispersion bends down. 
But this time refuses to saturate at a constant surface plasmon frequency. It linearly vanishes at some finite wave 
vector $q_{\rm ins}$ which signals an instability. The value of $q_{\rm ins}$ is independent of $\omega_b$ (or $|\bs b|$) 
and is only controlled by the tilt parameter $\zeta$. Furthermore in this situation, as shown in Fig.~\ref{fig13-yzz+1.fig}, 
a higher energy dispersive branch of SPP emerges the velocity of which is less than the light velocity, $c$. 

The situation with $b^\mu=(b_0,\bs 0)$ also features the formation of SPP-forbidden region 
for strong $\zeta$. However, if the tilt is in the $xy$ plane, the formation of SPP-forbidden region
requires stronger $\zeta$ when propagation direction $\bs q$ is transverse to $\bs\zeta$. 

\section{Acknowledgments}  S.A.J. was supported by research deputy of Sharif University of Technology, grant no. G960214 and the
Iran Science Elites Federation and the Max-Planck Institute for physics of complex systems, Dresden.

\bibliography{bib}

\end{document}